\documentclass[english,10pt,twocolumn,twoside]{IEEEtran}
\usepackage[T1]{fontenc}
\usepackage[latin9]{inputenc}
\usepackage{units}
\usepackage{bm}
\usepackage{amsthm}
\usepackage{amsmath}
\usepackage{amssymb}
\usepackage{graphicx}
\usepackage{esint}

\makeatletter
\theoremstyle{plain}

\theoremstyle{plain}
\newtheorem{prop}{\protect\propositionname}
\theoremstyle{plain}
\newtheorem{cor}{\protect\corollaryname}

\@ifundefined{showcaptionsetup}{}{%
 \PassOptionsToPackage{caption=false}{subfig}}
\usepackage{subfig}
\makeatother

\usepackage{babel}
\providecommand{\corollaryname}{Corollary}
\providecommand{\propositionname}{Proposition}
\providecommand{\theoremname}{Theorem}

\begin{document}

\title{Performance Analysis and Design of\\
 Maximum Ratio Combining in\\
 Channel-Aware MIMO Decision Fusion}

\author{Domenico~Ciuonzo,~\IEEEmembership{Student~Member,~IEEE,} Gianmarco~Romano,~\IEEEmembership{Member,~IEEE,}\\
 and Pierluigi~Salvo~Rossi,~\IEEEmembership{Senior Member,~IEEE}%
\thanks{The authors are with the Department of Industrial and Information
Engineering, Second University of Naples, Aversa (CE), Italy. Email:
\texttt{\{domenico.ciuonzo, gianmarco.romano, pierluigi.salvorossi\}@unina2.it}.%
}}
\maketitle
\begin{abstract}
In this paper we present a theoretical performance analysis of the
maximum ratio combining (MRC) rule for channel-aware decision fusion
over multiple-input multiple-output (MIMO) channels for (conditionally)
dependent and independent local decisions. The system probabilities
of false alarm and detection conditioned on the channel realization
are derived in closed form and an approximated threshold choice is
given. Furthermore, the channel-averaged (CA) performances are evaluated
in terms of the CA system probabilities of false alarm and detection
and the area under the receiver operating characteristic (ROC) through
the closed form of the conditional moment generating function (MGF)
of the MRC statistic, along with Gauss-Chebyshev (GC) quadrature rules.
Furthermore, we derive the deflection coefficients in closed form,
which are used for sensor threshold design. Finally, all the results
are confirmed through Monte Carlo simulations.\end{abstract}

\section{Introduction}

\subsection{Motivation and Related Works}

\PARstart{O}{ptimum} channel-aware decision fusion (DF) in wireless
sensor networks (WSNs) with instantaneous channel-state information
(CSI) is a challenging task and, due to the numerical instability
and strong requirements on the system knowledge required by the log-likelihood
ratio (LLR) test, several sub-optimal alternatives have been analyzed
in the recent literature, such as maximum ratio combining (MRC), equal
gain combining and Chair-Varshney maximum likelihood \cite{Chen2004,Chen2006}.
Max-Log rule has been studied in \cite{Lei2010} and shown to outperform
other sub-optimal rules though exhibiting higher complexity and requirements
on system knowledge, e.g. the channel variance. All the mentioned
rules were derived in the parallel-access channel (PAC) scenario and,
for such a case, a theoretical performance analysis was also conducted
in \cite{Lei2010}. 

Recently, DF exploiting the interfering nature of the broadcast wireless
medium is becoming more attractive for spectral-efficiency purposes.
Distributed detection over a multiple-access channel (MAC) is studied
in \cite{Li2007}, where perfect compensation of the fading coefficients
is assumed for each sensor. Non-coherent modulation and censoring
over PAC and MAC have been analyzed in \cite{Berger2009} with emphasis
on processing gain and combining loss. The same scenario is studied
in \cite{Li2011}, with a focus on the error exponents (obtained through
the large deviation principle) and the design of energy-efficient
modulations for Rayleigh and Rice fading. Optimality of received-energy
statistic in Rayleigh fading scenario is demonstrated for a diversity
MAC with non-identical sensors in \cite{Ciuonzo2013}. Efficient DF
over MAC only with knowledge of the instantaneous channel gains and
with the help of power-control and phase-shifting techniques is studied
in \cite{Umebayashi2012}. Techniques borrowed from direct-sequence
spread-spectrum systems are combined with on-off keying (OOK) modulation
and censoring for DF in scenarios with statistical CSI \cite{Yiu2009}.

DF with a multiple-input multiple-output (MIMO) wireless channel model
has been first studied in \cite{Zhang2008}, with a focus on power-allocation
design based on instantaneous CSI, under the framework of J-divergence.
Distributed detection with ultra-wideband sensors over MAC has been
then studied in \cite{Bai2010}; the same model was adopted to study
data fusion over MIMO channels with amplify and forward sensors in
\cite{Banavar2012}. 

Various sub-optimal fusion rules (with reduced system knowledge) for
channel-aware DF in the MIMO scenario with instantaneous CSI have
been proposed in \cite{Ciuonzo2012}, where decode-and-fuse and decode-then-fuse
approaches are compared through simulation results. It is worth noticing
that in such scenario the LLR is not a viable solution, since it suffers
from the exponential growth of the computational complexity with respect
to (w.r.t.) the number of sensors and high required system knowledge.

Differently, it has been shown that the MRC (sub-optimal) fusion rule
in MIMO scenario has the following appealing properties \cite{Ciuonzo2012}:
($i$) it exploits efficiently diversity from multiple antennas; ($ii$)
it achieves optimality at low signal-to-noise ratio (SNR); ($iii$)
its complexity is linear w.r.t. the number of antennas and independent
of the number of sensors, both in the fusion and channel estimation
stages; $(iv)$ it requires only limited system knowledge. Unfortunately,
MRC performance was (partially) assessed by relying on time-consuming
Monte Carlo (MC) simulations. Also, $(i)$ no explicit formula for
the choice of the threshold for the MRC fusion rule, assuring a given
false-alarm rate, was derived, ($ii$) no theoretical results on the
dependence of such fusion rule w.r.t. the WSN parameters (i.e. the
local-sensor performance, the channel SNR, the number of sensors and
the number of antennas) were presented and ($iii$) an asymptotic
analysis w.r.t. the mentioned parameters was lacking.

\subsection{Main Results and Paper Organization}

The main contributions of this manuscript are related to the MRC fusion
rule over MIMO channel and are summarized as follows:
\begin{itemize}
\item We obtain the closed form expressions of the instantaneous-channel
(IC) system probabilities of false alarm and detection, that are exploited
to: $(i)$ derive an approximate expression for the system threshold
in order to approach a target false-alarm rate, under \emph{low-SNR
and} \emph{large-system }(i.e. a high number of sensors) regime; ($ii$)
evaluate the IC system probabilities of false alarm and detection
under a \emph{large antenna array }(i.e. a high number of antennas)
regime at the DF center (DFC).
\item We derive the closed form expression of the channel-averaged (CA)
conditional moment generating function (MGF) of the statistic, i.e.
averaged over the statistical distribution of the channel; such a
result is used in conjunction with Gauss-Chebyshev (GC) quadrature
rules to \emph{efficiently} evaluate the CA probability of detection
and false alarm, as opposed to the results obtained via (time consuming)
MC simulations in \cite{Ciuonzo2012}; the obtained result is very
general, as it holds for scenarios with both (conditionally) dependent
and independent local decisions. 
\item The CA conditional MGF is exploited to show that large-system limit
under both individual power constraint (IPC) and total power constraint
(TPC) scenarios \emph{leads to a non-ideal receiver operating characteristic
}(ROC) in both cases. Such a result allows: ($i$) to claim zero error-exponents;
($ii$) a convenient evaluation of maximum achievable performance
with a fixed number of antennas (and a fixed SNR under TPC).
\item The CA conditional MGF is used in a new general formula for GC-based
computation of the \emph{area under the ROC }(AUC), thus allowing
synthetic and global performance analysis of the statistic.
\item Finally, we derive explicitly the (modified) deflection coefficient
\cite{Picinbono1995}, which is exploited in order to obtain a convenient
choice of the local threshold; the effectiveness of this approach
is confirmed via simulations.
\end{itemize}
The manuscript is organized as follows: Section \ref{sec:System-Model}
introduces the system model, while in Section \ref{sec:Fusion-Rules}
we recall the LLR and the MRC rules for the model under investigation;
in Section \ref{sec:MRC-Analytical-Resultsperformance} we present
the theoretical results needed for the performance analysis of MRC,
while some guidelines on the system design with MRC are obtained in
Section \ref{sec:System Design - Def. Coefficients}, via the deflection
coefficients; the results are verified and analyzed in Section \ref{sec:Simulations-Results};
finally in Section \ref{sec:Conclusions} we draw some concluding
remarks; proofs and derivations are confined to the Appendices.

\emph{Notation} - Lower-case (resp. Upper-case) bold letters denote
vectors (resp. matrices), with $a_{n}$ (resp. $a_{n,m}$) representing
the $n$th (resp. the $(n,m)$th) element of $\bm{a}$ (resp. $\bm{A}$);
upper-case calligraphic letters denote finite sets, with $\mathcal{A}^{K}$
representing the $k$-ary Cartesian power of $\mathcal{A}$; $\bm{I}_{N}$
denotes the $N\times N$ identity matrix; $\bm{0}_{N}$ (resp. $\bm{1}_{N}$)
denotes the null (resp. ones) vector of length $N$; $\mathbb{E}\{\cdot\}$,
$\mathrm{var\{\cdot\}}$, $(\cdot)^{t}$, $(\cdot)^{\dagger}$, $\Re\left(\cdot\right)$,
$\Im(\cdot)$, $\left\Vert \cdot\right\Vert $ and $\mathrm{det}(\cdot)$
denote expectation, variance, transpose, conjugate transpose, real
part, imaginary part, Frobenius norm and matrix determinant operators,
respectively; $j$ denotes the imaginary unit; $P(\cdot)$ and $p(\cdot)$
denote probability mass functions (pmf) and probability density functions
(pdf), while $P(\cdot|\cdot)$ and $p(\cdot|\cdot)$ their corresponding
conditional counterparts; $\mathcal{N}_{\mathbb{C}}(\bm{\mu},\bm{\Sigma})$
(resp. $\mathcal{N}(\bm{\mu},\bm{\Sigma})$) denotes a circularly
symmetric complex (resp. real) normal distribution with mean vector
$\bm{\mu}$ and covariance matrix $\bm{\Sigma}$, while $\mathcal{Q}(\cdot)$
is the complementary cumulative distribution function of a standard
normal distribution; $\mathcal{B}(k,p)$ denotes a binomial distribution
of $k$ trials with probability of success $p$; $\Gamma(k,\theta)$
denotes a Gamma distribution with shape parameter $k$ and scale parameter
$\theta$; finally the symbols $\propto$, $\rightarrow$, $\overset{d}{\rightarrow}$
and $\sim$ mean ``proportional to'', ``tends to'', ``tends in
distribution to'' and \textquotedblleft{}distributed as\textquotedblright{},
respectively.

\section{System Model\label{sec:System-Model}}

\begin{figure}
\begin{centering}
\includegraphics[width=1\columnwidth]{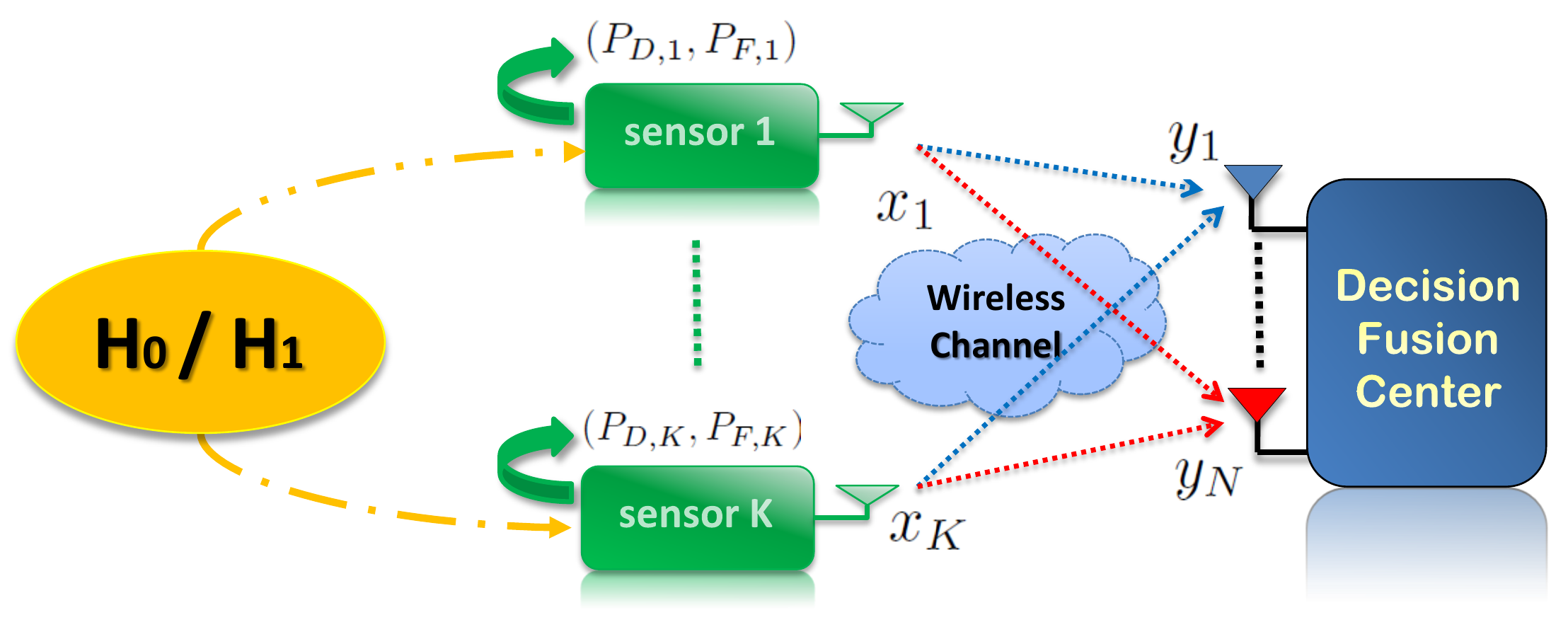}
\par\end{centering}

\caption{The DF model in presence of a (virtual) MIMO channel.\label{fig:system model}}
\end{figure}

In this section we briefly describe the system model, illustrated
in Fig. \ref{fig:system model}. We consider a distributed binary
hypothesis test, where $K$ sensors are used to discriminate between
the hypotheses of the set $\mathcal{H}\triangleq\{\mathcal{H}_{0},\mathcal{H}_{1}\}$.
For example $\mathcal{H}_{0}$ and $\mathcal{H}_{1}$ may represent
the absence and the presence of a specific target of interest, respectively.
The $k$th sensor, $k\in\mathcal{K}\triangleq\{1,2,\ldots,K\}$, takes
a binary local decision $d_{k}\in\mathcal{H}$ about the observed
phenomenon on the basis of its own measurements. Here \emph{we do
not make} \emph{any conditional} (given $\mathcal{H}_{i}\in\mathcal{H}$)
\emph{mutual independence assumption} on $d_{k}$. Each decision $d_{k}$
is mapped to a symbol $x_{k}\in{\cal X}=\{-1,+1\}$ representing a
binary phase-shift keying (BPSK) modulation%
\footnote{In the case of an absence/presence task, where ${\cal H}_{0}$ is
much more probable, OOK can be employed for energy-efficiency purposes.
Hereinafter we will refer only to BPSK, however the results apply
readily to OOK.%
}: without loss of generality (w.l.o.g.) we assume that $d_{k}=\mathcal{H}_{i}$
maps into $x_{k}=2i-1$, $i\in\{0,1\}$. The quality of the $k$th
sensor decisions is characterized by the conditional probabilities
$P(x_{k}|\mathcal{H}_{i})$. More specifically, we denote $P_{D,k}\triangleq P\left(x_{k}=1|\mathcal{H}_{1}\right)$
and $P_{F,k}\triangleq P\left(x_{k}=1|\mathcal{H}_{0}\right)$ the
probability of detection and false alarm of the $k$th sensor, respectively.

Sensors communicate with DFC over a wireless flat-fading MAC, with
independent and identically distributed (i.i.d.) Rayleigh fading coefficients
of unitary mean power. The DFC is equipped with $N$ receive antennas
in order to exploit diversity and combat signal attenuation due to
small-scale fading; this configuration determines a distributed or
``virtual'' MIMO\textbf{ }channel \cite{Zhang2008,Ciuonzo2012}.
Also, instantaneous CSI and perfect synchronization%
\footnote{Multiple antennas at the DFC do not make these assumptions harder
to verify w.r.t. a single-antenna MAC.%
} are assumed at the DFC as in \cite{Li2007,Ciuonzo2013,Zhang2008,Ciuonzo2012}. 

We denote: $y_{n}$ the received signal at the $n$th receive antenna
of the DFC after matched filtering and sampling; $h_{n,k}\sim{\cal N}_{\mathbb{C}}\left(0,1\right)$
the fading coefficient between the $k$th sensor and the $n$th receive
antenna of the DFC; $w_{n}$ the additive white Gaussian noise at
the $n$th receive antenna of the DFC. The vector model at the DFC
is the following:
\begin{equation}
\bm{y}=\bm{H}\bm{x}+\bm{w}\label{eq:channel_model}
\end{equation}
where $\bm{y}\in\mathbb{C}^{N}$, $\bm{H}\in\mathbb{C}^{N\times K}$,
$\bm{x}\in\mathcal{X}^{K}$, $\bm{w}\sim\mathcal{N}_{\mathbb{C}}(\bm{0}_{N},\sigma_{w}^{2}\bm{I}_{N})$
are the received-signal vector, the channel matrix, the transmitted-signal
vector and the noise vector, respectively. Furthermore, we define
the random variable (r.v.) $\ell(\bm{x})$ denoting the number of
sensors deciding $\mathcal{H}_{1}$ and the set $\mathcal{L}\triangleq\{0,\ldots,K\}$,
denoting the outcomes of $\ell(\bm{x})$.

Finally, we define the total channel SNR as the ratio between the
energy transmitted from the WSN $\mathcal{E}_{s}\triangleq\mathbb{E}\{\left\Vert \bm{x}\right\Vert {}^{2}\}=K$
and the one-sided power spectral density of the noise $\sigma_{w}^{2}$,
i.e. $\mathrm{SNR}\triangleq\frac{\varepsilon_{S}}{\sigma_{w}^{2}}=\frac{K}{\sigma_{w}^{2}}$;
therefore the individual channel SNR for the $k$th sensor will be
$\mathrm{SNR}_{\star}=\frac{1}{\sigma_{w}^{2}}$. Hereinafter we will
consider in our analysis both IPC (i.e. fixed $\mathrm{SNR}_{\star}$)
and TPC (i.e. fixed $\mathrm{SNR}$) on the WSN.

\section{Fusion Rules\label{sec:Fusion-Rules}}

\subsection{Optimum Rule}

The optimal test \cite{Kay1998} for the considered problem is formulated
as 
\begin{equation}
\Lambda_{opt}\triangleq\ln\left[\frac{p(\bm{y}|\bm{H},\mathcal{H}_{1})}{p(\bm{y}|\bm{H},\mathcal{H}_{0})}\right]\begin{array}{c}
{\scriptstyle \hat{\mathcal{H}}=\mathcal{H}_{1}}\\
\gtrless\\
{\scriptstyle \hat{\mathcal{H}}=\mathcal{H}_{0}}
\end{array}\gamma\label{eq:neyman_pearson_test}
\end{equation}
where $\hat{\mathcal{H}}$, $\Lambda_{opt}$ and $\gamma$ denote
the estimated hypothesis, the LLR and the threshold which the LLR
is compared to. The threshold $\gamma$ can be determined to assure
a fixed system false-alarm rate (Neyman-Pearson approach), or can
be chosen to minimize the probability of error (Bayesian approach)
\cite{Kay1998}. Exploiting the independence%
\footnote{In fact, as shown in Fig. \ref{fig:system model}, the directed triple
formed by hypothesis, the transmitted-signal vector and the received-signal
vector satisfies the Markov property.%
} of $\bm{y}$ from $\mathcal{H}_{i}$, given $\bm{x}$, an explicit
expression of the LLR from Eq.~(\ref{eq:neyman_pearson_test}) is
given by 
\begin{align}
\Lambda_{opt} & =\ln\left[\frac{\sum_{\bm{x}\in{\cal X}^{K}}p(\bm{y}|\bm{H},\bm{x})P(\bm{x}|\mathcal{H}_{1})}{\sum_{\bm{x}\in{\cal X}^{K}}p(\bm{y}|\bm{H},\bm{x})P(\bm{x}|\mathcal{H}_{0})}\right]\label{eq:optimum_llr}\\
 & =\ln\left[\frac{\sum_{\bm{x}\in{\cal X}^{K}}\exp\left(-\frac{\bm{\|y}-\bm{H}\bm{x}\|^{2}}{\sigma_{w}^{2}}\right)P(\bm{x}|\mathcal{H}_{1})}{\sum_{\bm{x}\in{\cal X}^{K}}\exp\left(-\frac{\bm{\|y}-\bm{H}\bm{x}\|^{2}}{\sigma_{w}^{2}}\right)P(\bm{x}|\mathcal{H}_{0})}\right].\nonumber 
\end{align}

Unfortunately, the optimal rule in Eq. (\ref{eq:optimum_llr}) presents
several difficulties in the implementation: ($i$) complete knowledge
of $\bm{H}$, $P(\bm{x}|\mathcal{H}_{i})$ and $\sigma_{w}^{2}$;
($ii$) numerical instability of the expression, due to the presence
of exponential functions with large dynamics \cite{Lei2010,Ciuonzo2012};
($iii$) exponential growth of the complexity with $K$. Design of
sub-optimal DF rules with simpler implementation and reduced system
knowledge is then extremely desirable.

\subsection{MRC}

The LLR of Eq.~(\ref{eq:optimum_llr}) can be simplified under the
assumption of perfect sensors \cite{Lei2010,Yiu2009}, i.e. $P(\bm{x}=\bm{1}_{K}|\mathcal{H}_{1})=P(\bm{x}=-\bm{1}_{K}|\mathcal{H}_{0})=1$.
In this case $\bm{x}\in\{\bm{1}_{K},-\bm{1}_{K}\}$ and Eq.~(\ref{eq:optimum_llr})
reduces to \cite{Ciuonzo2012}:
\begin{eqnarray}
\ln\left[\frac{\exp\left(-\frac{\|\bm{y}-\bm{H}\bm{1}_{K}\|^{2}}{\sigma_{w}^{2}}\right)}{\exp\left(-\frac{\|\bm{y}+\bm{H}\bm{1}_{K}\|^{2}}{\sigma_{w}^{2}}\right)}\right] & \propto & \Re(\bm{z}_{{\scriptscriptstyle MRC}}^{\dagger}\bm{y})\triangleq\Lambda_{{\scriptscriptstyle MRC}}\label{eq:MRC}\\
{\rm with} &  & \bm{z}_{{\scriptscriptstyle MRC}}\triangleq\bm{H}\bm{1}_{K}\nonumber 
\end{eqnarray}
where terms independent on $\bm{y}$ have been incorporated in $\gamma$
as in Eq.~(\ref{eq:neyman_pearson_test}). It is worth noticing that
the MRC is a \emph{sub-optimal} rule since, in the practice, the sensor
local decisions are far from being perfect\emph{. }However, it has
been proved in \cite{Ciuonzo2012} that MRC is the low-$\mathrm{SNR}$
approximation of the optimum of Eq.~(\ref{eq:optimum_llr}) when
local performances of sensors are identical%
\footnote{Even if in \cite{Ciuonzo2012} conditional mutual independence of
local decisions was assumed, it can be shown by inspection of the
derivation that such an assumption is not necessary in proving MRC
optimality at low-$\mathrm{SNR}$.%
}. Furthermore, as stated in the introduction, its computational complexity
is independent of $K$ in both fusion and channel-estimation stages,
since $\bm{z}_{{\scriptscriptstyle MRC}}=\bm{H}\bm{1}_{K}$ (which
is the only required parameter for the rule implementation, as opposed
to the LLR) can be directly estimated%
\footnote{This is easily obtained with a channel estimation procedure in which
all the sensors transmit $x_{k}=1$ (or equivalently $x_{k}=-1$).
However, time-varying channels in high-mobility scenarios may be problematic
as the channel state information could be outdated when available
\cite{Biguesh2006}.%
} and used in Eq. (\ref{eq:MRC}).

\section{MRC performance analysis\label{sec:MRC-Analytical-Resultsperformance}}

\subsection{IC False Alarm and Detection probabilities and threshold computation}

The IC system probabilities of false alarm and detection are defined
as 
\begin{gather}
P_{F_{0}}(\bm{H},\gamma)\triangleq P(\Lambda>\gamma|\bm{H},\mathcal{H}_{0}),\label{eq:ic_PD0}\\
P_{D_{0}}(\bm{H},\gamma)\triangleq P(\Lambda>\gamma|\bm{H},\mathcal{H}_{1}),\label{eq:ic_PF0}
\end{gather}
with $\Lambda$ representing the decision statistic of a generic fusion
rule. It is shown in Appendix \ref{sec:Appendix_Proposition-threshold-lowsnr}
that
\begin{gather}
\Lambda_{{\scriptscriptstyle MRC}}|\bm{H},\mathcal{H}_{i}\sim\nonumber \\
\sum_{\bm{x}\in\mathcal{X}^{K}}P(\bm{x}|\mathcal{H}_{i})\mathcal{N}\left(\Re(\bm{z}_{{\scriptscriptstyle MRC}}^{\dagger}\bm{H}\bm{x}),\frac{\sigma_{w}^{2}}{2}\left\Vert \bm{z}_{{\scriptscriptstyle MRC}}\right\Vert ^{2}\right),\label{eq:istc_pdf_MRC}
\end{gather}
i.e. the pdf $p_{\Lambda_{MRC}}(\lambda|\bm{H},\mathcal{H}_{i})$,
$\lambda\in\mathbb{R}$, $\mathcal{H}_{i}\in\mathcal{H}$, is a \emph{Gaussian
mixture}. Therefore, combining Eqs. (\ref{eq:ic_PD0}), (\ref{eq:ic_PF0})
and (\ref{eq:istc_pdf_MRC}), leads to:
\begin{align}
P_{F_{0}}(\bm{H},\gamma) & =\sum_{\bm{x}\in\mathcal{X}^{K}}P(\bm{x}|\mathcal{H}_{0})\mathcal{Q}\left(\frac{\gamma-\Re(\bm{z}_{{\scriptscriptstyle MRC}}^{\dagger}\bm{H}\bm{x})}{\sqrt{\frac{1}{2}}\sigma_{w}\left\Vert \bm{z}_{{\scriptscriptstyle MRC}}\right\Vert }\right);\label{eq:ic_PF0_closedform}\\
P_{D_{0}}(\bm{H},\gamma) & =\sum_{\bm{x}\in\mathcal{X}^{K}}P(\bm{x}|\mathcal{H}_{1})\mathcal{Q}\left(\frac{\gamma-\Re(\bm{z}_{{\scriptscriptstyle MRC}}^{\dagger}\bm{H}\bm{x})}{\sqrt{\frac{1}{2}}\sigma_{w}\left\Vert \bm{z}_{{\scriptscriptstyle MRC}}\right\Vert }\right).\label{eq:ic_PD0_closedform}
\end{align}
The explicit expression of $P_{F_{0}}(\bm{H})$ in Eq. (\ref{eq:ic_PF0_closedform})
cannot be easily used to design a threshold $\gamma$ which satisfies
a given false-alarm rate, since a two-fold complication is present:
($i$) the inversion can be done only numerically and requires the
evaluation of a sum over $2^{K}$ terms; ($ii$) complete knowledge
of $\bm{H}$ (and not of only $\bm{z}_{{\scriptscriptstyle MRC}}=\bm{H}\bm{1}_{K}$,
as required instead from the MRC rule), $P(\bm{x}|\mathcal{H}_{0})$
and $\sigma_{w}^{2}$ is required. 

Nonetheless, a low-SNR large-system ($K\rightarrow+\infty$) approximation
of $\gamma$ with reduced system knowledge can be found, given a target
$\breve{P}_{F_{0}}$; the result is stated by the following proposition,
in the case of (conditionally) uncorrelated sensor decisions (under
$\mathcal{H}_{0}$) and $P_{F,k}=P_{F}$, $k\in\mathcal{K}$.
\begin{prop}
Assuming ($i$) $\mathbb{E}\{\bm{x}|\mathcal{H}_{0}\}=\bm{\mu}_{0}\triangleq(2P_{F}-1)\bm{1}_{K}$
and ($ii$) $\mathbb{E}\{(\bm{x}-\bm{\mu}_{0})(\bm{x}-\bm{\mu}_{0})^{T}|\mathcal{H}_{0}\}=\left[1-(2P_{F}-1)^{2}\right]\cdot\bm{I}_{K}$,
a low-$\mathrm{SNR}$ large-system $\breve{\gamma}$ for approaching
a target $\breve{P}_{F_{0}}$, is given by\label{prop:threshold low-SNR}
\begin{align}
\breve{\gamma} & \triangleq\mathcal{Q}^{-1}(\breve{P}_{F_{0}})\sqrt{\frac{(1-\delta^{2})K+\sigma_{w}^{2}}{2}}\left\Vert \bm{z}_{{\scriptscriptstyle MRC}}\right\Vert +\delta\left\Vert \bm{z}_{{\scriptscriptstyle MRC}}\right\Vert ^{2}\label{eq:low-SNR large-system threshold}
\end{align}
where $\delta\triangleq(2P_{F}-1)$.\end{prop}
\begin{IEEEproof}
The proof is given in Appendix \ref{sec:Appendix_Proposition-threshold-lowsnr}.
\end{IEEEproof}
The accuracy of Eq. (\ref{eq:low-SNR large-system threshold}) will
be verified in Section~\ref{sec:Simulations-Results}; it is worth
noticing that such expression \emph{does not require} the complete
knowledge of $\bm{H}$ (the dependence is only through $\bm{z}_{{\scriptscriptstyle MRC}}=\bm{H}\bm{1}_{K}$)
and $P(\bm{x}|\mathcal{H}_{0})$. Also, the assumptions on $P(\bm{x}|\mathcal{H}_{0})$
in Proposition \ref{prop:threshold low-SNR} are generally verified
when the local threshold at each sensor is set up to satisfy the same
false-alarm rate (assumption ($i$)) and the local decisions are \emph{uncorrelated
}under the hypothesis $\mathcal{H}_{0}$ (assumption ($ii$)), which
is typically the case when $\mathcal{H}_{0}$ corresponds to the absence
of an event of interest.

Furthermore, it can be shown that for large $N$ at the DFC, Eq.~(\ref{eq:istc_pdf_MRC})
reduces to
\begin{gather}
\Lambda{}_{{\scriptscriptstyle MRC}}|\bm{H},\mathcal{H}_{i}\overset{{\scriptscriptstyle approx.}}{\sim}\nonumber \\
\sum_{\bm{x}\in\mathcal{X}^{K}}P(\bm{x}|\mathcal{H}_{i})\mathcal{N}\left(\left(2\ell(\bm{x})-K\right)N,\frac{\sigma_{w}^{2}}{2}NK\right)=\\
\sum_{\ell(\bm{x})=0}^{K}P(\ell(\bm{x})|\mathcal{H}_{i})\mathcal{N}\left(\left(2\ell(\bm{x})-K\right)N,\frac{\sigma_{w}^{2}}{2}NK\right),
\end{gather}
since, when $N$ is large, the approximation $\bm{H}^{\dagger}\bm{H}\approx N\bm{I}_{K}$
holds \cite{Marzetta2010}. It is worth noticing that a large antenna
array at the DFC on one hand makes the performance independent of
the particular instance of $\bm{H}$, on the other hand it ``reduces''
the dependence of the MRC performances w.r.t. the (joint) sensor performance,
i.e. requires only $P(\ell(\bm{x})|\mathcal{H}_{i})$ as opposed to
$P(\bm{x}|\mathcal{H}_{i}$) (cf. Eq. (\ref{eq:istc_pdf_MRC})). This
result is confirmed by observing that, for large $N$, Eq. (\ref{eq:MRC})
reduces to 
\begin{equation}
\Lambda_{{\scriptscriptstyle MRC}}\approx N\cdot\sum_{k=1}^{K}x_{k}+\Re\{\bar{w}\},\label{eq:large antenna array MRC}
\end{equation}
where $\bar{w}\sim\mathcal{N}_{\mathbb{C}}(0,\sigma_{w}^{2}NK)$,
i.e. the MRC approaches a ``noisy'' counting rule \cite{Varshney1996}.

\subsection{CA False Alarm and Detection probabilities }

The CA system probabilities of false alarm and detection are 
\begin{gather}
P_{F_{0}}(\gamma)\triangleq\mathbb{E}_{\bm{H}}\{P_{F_{0}}(\gamma,\bm{H})\}=P(-\Lambda<-\gamma|\mathcal{H}_{0}),\label{eq:CA_PF_0}\\
P_{D_{0}}(\gamma)\triangleq\mathbb{E}_{\bm{H}}\{P_{D_{0}}(\gamma,\bm{H})\}=P(-\Lambda<-\gamma|\mathcal{H}_{1}),\label{eq:CA_PD_0}
\end{gather}
with $\Lambda$ representing the decision statistic of a generic fusion
rule. It is worth noticing that Eqs. (\ref{eq:CA_PF_0}) and (\ref{eq:CA_PD_0})
are formulated in terms of $-\Lambda$ in order to exploit readily
the standard definition of the conditional MGFs in the Laplace domain
\cite{Lei2010}. Although it is often difficult to derive the conditional
pdf $p_{-\Lambda}(\lambda|\mathcal{H}_{i})$, $\lambda\in\mathbb{R}$,
$\mathcal{H}_{i}\in\mathcal{H}$, of the r.v. $-\Lambda$, the corresponding
Laplace transform $\Phi_{-\Lambda}(s|\mathcal{H}_{i})$ (i.e. the
MGF of $\Lambda|\mathcal{H}_{i}$) is usually easier to obtain. Using
the relationship $p_{-\Lambda}(\lambda|\mathcal{H}_{i})=\frac{1}{2\pi j}\int_{c-j\infty}^{c+j\infty}\Phi_{-\Lambda}(s|\mathcal{H}_{i})\exp\left(\lambda s\right)ds$,
where $c$ is a small (positive) constant in the region of convergence
(RC) of the integral, both probabilities in Eqs. (\ref{eq:CA_PF_0})
and (\ref{eq:CA_PD_0}) can be rewritten as
\begin{align}
\intop_{-\infty}^{-\gamma}p_{-\Lambda}(\lambda|\mathcal{H}_{i})d\lambda= & \intop_{c-j\infty}^{c+j\infty}\Phi_{-\Lambda}(s|\mathcal{H}_{i})\,\frac{\exp\left(-\gamma s\right)}{2\pi j}\,\frac{ds}{s}.\label{eq:laplace_integral}
\end{align}

Based on Eq. (\ref{eq:laplace_integral}), $P_{F_{0}}(\gamma)$ and
$P_{D_{0}}(\gamma)$ can be calculated for any fusion rule provided
that the integral in Eq.~(\ref{eq:laplace_integral}) can be solved
efficiently and the corresponding Laplace transform $\Phi_{-\Lambda}(s|\mathcal{H}_{i})$
can be derived in closed form. It is worth remarking that the same
approach was used to efficiently evaluate CA probabilities of sub-optimal
fusion rules over PAC in \cite{Lei2010}. 

The integral in Eq.~(\ref{eq:laplace_integral}) can be solved exactly
using the residue approach or numerically through GC quadrature rules
\cite{Biglieri1996,Annamalai1998,Biglieri1998}. Unfortunately, the
former approach becomes long and intricate when poles of algebraic
multiplicity greater than one are present (indeed this is our case,
since we are considering multiple antennas at the DFC) \cite{Biglieri1998}.
On the other hand, following the latter approach, a direct application
of the results in \cite{Biglieri1996} to Eq. (\ref{eq:laplace_integral})
leads to
\begin{align}
\intop_{-\infty}^{-\gamma}p_{-\Lambda}(\lambda|\mathcal{H}_{i})d\lambda & \approx\frac{1}{\nu}\sum_{r=1}^{\nu/2}[\Re\left(\varphi_{i}(r)\right)+\tau_{r}\cdot\Im\left(\varphi_{i}(r)\right)],\label{eq:GC explicit formula_Pd0-Pf0}\\
\varphi_{i}(r) & \triangleq\Phi_{-\Lambda}(\mu_{r}|\mathcal{H}_{i})\exp\left(-\gamma\cdot\mu_{r}\right),\label{eq:Complex Function quadrature Pd0-Pf0}
\end{align}
 where $\nu$ denotes the (even) \emph{number of nodes} of the GC
rules (i.e. the order of the approximation accuracy), $\mu_{r}\triangleq c+jc\tau_{r}$
and $\tau_{r}\triangleq\tan\left(\frac{(2r-1)\pi}{2\nu}\right)$. 

Differently, given the assumptions of the model under consideration,
$\Phi_{-\Lambda}(s|\mathcal{H}_{i})$ can be expanded as 
\begin{equation}
\Phi_{-\Lambda}(s|\mathcal{H}_{i})=\sum_{\bm{x}\in\mathcal{X}^{K}}\Phi_{-\Lambda}(s|\bm{x})P(\bm{x}|\mathcal{H}_{i}),\qquad\mathcal{H}_{i}\in\mathcal{H}.\label{eq:MGFexpansion}
\end{equation}
We derive here $\Phi_{-\Lambda_{MRC}}(s|\bm{x})$ in closed form,
as summarized by the following proposition.
\begin{prop}
The Laplace Transform $\Phi_{-\Lambda_{MRC}}(s|\bm{x})$ of $p_{-\Lambda_{{\scriptscriptstyle MRC}}}(\lambda|\bm{x})$
is given in closed form in\label{prop:MGF_mrc} Eq. (\ref{eq:mgf_lambda_mrc})
at the top of the next page, where $\xi^{+}\triangleq1+\sqrt{1+\frac{1}{\mathrm{SNR}}}$
and $\xi^{-}\triangleq1-\sqrt{1+\frac{1}{\mathrm{SNR}}}$.
\end{prop}
\begin{figure*}[t]
\begin{equation}
\Phi_{-\Lambda_{MRC}}(s|\bm{x})=\frac{1}{\left\{ \left[1+\frac{1}{2}\left(K\cdot\xi^{+}-2\ell(\bm{x})\right)s\right]\left[1+\frac{1}{2}\left(K\cdot\xi^{-}-2\ell(\bm{x})\right)s\right]\right\} ^{N}}.\label{eq:mgf_lambda_mrc}
\end{equation}

% IEEE uses as a separator 
\hrulefill 
% The spacer can be tweaked to stop underfull vboxes. 
\vspace*{0pt}
\end{figure*}

\begin{IEEEproof}
The proof is given in Appendix \ref{sec:Appendix_MGF-MRC}.
\end{IEEEproof}
It is worth noticing that, in the particular case of the MRC rule,
Eq. (\ref{eq:mgf_lambda_mrc}) depends on $\bm{x}$ only through $\ell(\bm{x})$,
i.e. $\Phi_{-\Lambda_{MRC}}(s|\bm{x})=\Phi_{-\Lambda_{MRC}}(s|\ell(\bm{x}))$.
Then Eq.~(\ref{eq:MGFexpansion}) is replaced efficiently with
\begin{equation}
\Phi_{-\Lambda_{MRC}}(s|\mathcal{H}_{i})=\sum_{\ell(\bm{x})=0}^{K}\Phi_{-\Lambda_{MRC}}(s|\ell(\bm{x}))P(\ell(\bm{x})|\mathcal{H}_{i}),\label{eq:MGF_efficient_expansion}
\end{equation}
requiring only a sum over $\left(K+1\right)$ terms ($\ell(\bm{x})\in\mathcal{L}$),
as opposed to $2^{K}$ (cf. with Eq. (\ref{eq:MGFexpansion})). Also,
only $P(\ell(\bm{x})|\mathcal{H}_{i})$, in the place of $P(\bm{x}|\mathcal{H}_{i})$,
is needed to evaluate $\Phi_{-\Lambda_{MRC}}(s|\mathcal{H}_{i})$. 

\emph{Remarks:} from inspection of Eqs. (\ref{eq:mgf_lambda_mrc})
and (\ref{eq:MGF_efficient_expansion}), it can be shown that the
RC of $\Phi_{-\Lambda_{MRC}}(s|\mathcal{H}_{i})$ is a vertical strip
delimited by the axes determined by $\Re(s)=\pm\frac{2}{K+K\sqrt{1+\frac{1}{\mathrm{SNR}}}}$.

\subsection{Large-system analysis\label{sub:Large System WSN analysis}}

Taking a closer look at Eqs. (\ref{eq:mgf_lambda_mrc}) and (\ref{eq:MGF_efficient_expansion})
the large-system ($K\rightarrow+\infty$) behaviour of the MRC under
both IPC and TPC is not apparent. Such behaviour is put in evidence
by the following proposition for the statistic%
\footnote{Note that considering $\tilde{\Lambda}$, in the place of $\Lambda_{{\scriptscriptstyle MRC}}$,
does not change MRC performance, since every positive constant can
be absorbed by the threshold $\gamma$ through Eq. (\ref{eq:neyman_pearson_test}).
Nonetheless, the scaling factor $\nicefrac{1}{K}$ is added in order
to assure convergence of the limit $K\rightarrow+\infty$ for the
MGFs being considered.%
} $\tilde{\Lambda}\triangleq\frac{\Lambda_{MRC}}{K},$ in the case
of conditionally i.i.d. sensor decisions.
\begin{prop}
If $P(\bm{x}|\mathcal{H}_{i})=\prod_{k=1}^{K}P(x_{k}|\mathcal{H}_{i})$,
$\mathcal{H}_{i}\in\mathcal{H},$ and $(P_{D,k},P_{F,k})=(P_{D},P_{F})$,
$k\in\mathcal{K}$, as $K\rightarrow+\infty$, the Laplace transform
$\Phi_{-\tilde{\Lambda}}(s|\mathcal{H}_{i})$ of $p_{-\tilde{\Lambda}}(\lambda|\mathcal{H}_{i})$,
in the IPC scenario is given by\label{prop:largesystemMGFs}:
\begin{align}
\bar{\Phi}_{-\tilde{\Lambda}}^{{\scriptscriptstyle I}}(s|\mathcal{H}_{1}) & =\frac{1}{\left\{ \left[1+\left(1-P_{D}\right)s\right]\left[1-P_{D}s\right]\right\} ^{N}};\label{eq:MGFH1_kinf_IPC}\\
\bar{\Phi}_{-\tilde{\Lambda}}^{{\scriptscriptstyle I}}(s|\mathcal{H}_{0}) & =\frac{1}{\left\{ \left[1+\left(1-P_{F}\right)s\right]\left[1-P_{F}s\right]\right\} ^{N}}.\label{eq:MGFH0_kinf_IPC}
\end{align}
Correspondingly, as $K\rightarrow+\infty$, $\Phi_{-\tilde{\Lambda}}(s|\mathcal{H}_{i})$
in the TPC scenario is given by:
\begin{gather}
\bar{\Phi}_{-\tilde{\Lambda}}^{{\scriptscriptstyle T}}(s|\mathcal{H}_{1})=\label{eq:MGFH1_kinf_TPC}\\
\frac{1}{\left\{ \left[1+\frac{1}{2}\left(\xi^{{\scriptscriptstyle +}}-2P_{D}\right)s\right]\left[1+\frac{1}{2}\left(\xi^{{\scriptscriptstyle -}}-2P_{D}\right)s\right]\right\} ^{N}}\nonumber \\
\bar{\Phi}_{-\tilde{\Lambda}}^{{\scriptscriptstyle T}}(s|\mathcal{H}_{0})=\label{eq:MGFH0_kinf_TPC}\\
\frac{1}{\left\{ \left[1+\frac{1}{2}\left(\xi^{{\scriptscriptstyle +}}-2P_{F}\right)s\right]\left[1+\frac{1}{2}\left(\xi^{{\scriptscriptstyle -}}-2P_{F}\right)s\right]\right\} ^{N}}\nonumber 
\end{gather}
\end{prop}
\begin{IEEEproof}
The proof is given in Appendix \ref{sec:Appendix_LargesystemMGF}.
\end{IEEEproof}
\emph{Remarks: }from inspection of Eq. (\ref{eq:MGFH1_kinf_IPC})
(resp. Eq. (\ref{eq:MGFH0_kinf_IPC})), it can be shown that the RC
of $\bar{\Phi}_{-\tilde{\Lambda}}^{{\scriptscriptstyle I}}(s|\mathcal{H}_{1})$
(resp. $\bar{\Phi}_{-\tilde{\Lambda}}^{{\scriptscriptstyle I}}(s|\mathcal{H}_{0})$)
is a vertical strip delimited by the axes determined by $\Re(s)=-\left(\frac{1}{1-P_{D}}\right)$
(resp. $\Re(s)=-\left(\frac{1}{1-P_{F}}\right)$) and $\Re(s)=\frac{1}{P_{D}}$
(resp. $\Re(s)=\frac{1}{P_{F}}$). Differently, $\bar{\Phi}_{-\tilde{\Lambda}}^{{\scriptscriptstyle T}}(s|\mathcal{H}_{1})$
(resp. $\bar{\Phi}_{-\tilde{\Lambda}}^{{\scriptscriptstyle T}}(s|\mathcal{H}_{0})$)
in Eq. (\ref{eq:MGFH1_kinf_TPC}) (resp. Eq. (\ref{eq:MGFH0_kinf_TPC}))
has a RC which is is a vertical-strip delimited by the axes determined
by $\Re(s)=\frac{2}{2P_{D}-\xi^{+}}$ (resp. $\Re(s)=\frac{2}{2P_{F}-\xi^{+}}$)
and $\Re(s)=\frac{2}{2P_{D}-\xi^{-}}$ (resp. $\Re(s)=\frac{2}{2P_{F}-\xi^{-}}$). 

In Fig. \ref{fig:Asymptotic-ROC-IPC TPC} it is illustrated the large-system
CA-ROC, i.e. $P_{D_{0}}$ vs $P_{F_{0}}$, obtained through the GC
rules (with $\nu=10^{3}$) for both IPC and TPC cases and several
configurations%
\footnote{Note that looking at the figure, the concavity of the ROCs is not
apparent, as instead suggested from the theory \cite{Kay1998}; this
is motivated by the use (throughout the paper) of a log-linear scale
in the plot.%
}. Some important considerations are expressed hereinafter:
\begin{figure}
\centering{}\includegraphics[width=0.95\columnwidth]{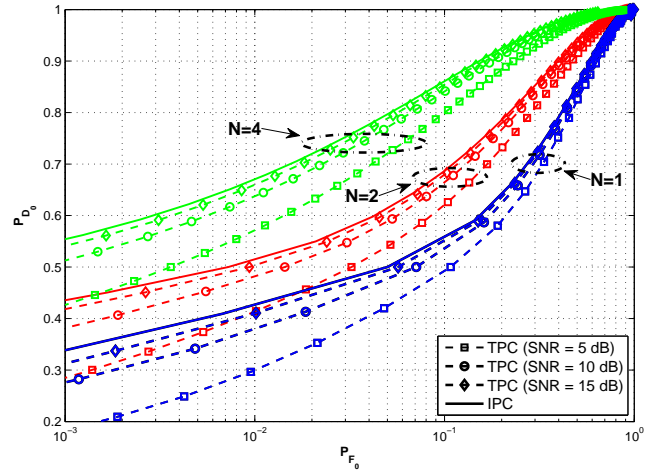}\caption{Large system ($K\rightarrow+\infty$) ROC for conditionally i.i.d.
sensor decisions in IPC and TPC case ($(\mathrm{SNR})_{dB}\in\{5,10,15\}$);
$(P_{D,k},P_{F,k})=(0.5,0.05)$, $k\in\mathcal{K}$.\label{fig:Asymptotic-ROC-IPC TPC}}
\end{figure}

\begin{itemize}
\item From inspection of Fig. \ref{fig:Asymptotic-ROC-IPC TPC}, it is apparent
that both in IPC and TPC cases the ROC \emph{can not be driven} toward
the point $(P_{D_{0},}P_{F_{0}})=(1,0)$ \emph{merely increasing the
number of sensors} $K$, as long as the number of antennas $N$ is
kept finite, thus leading to \emph{zero (Bayesian and Neyman-Pearson)
error-exponents} \cite{Dembo1993}. Such results are analogous to
the case of non-coherent DF with energy detection over diversity MAC
\cite{Ciuonzo2013}.
\item It is worth noticing that $\bar{\Phi}_{-\tilde{\Lambda}}^{{\scriptscriptstyle I}}(s|\mathcal{H}_{i})$
does not depend on $\mathrm{SNR}_{\star}$, as opposed to $\bar{\Phi}_{-\tilde{\Lambda}}^{{\scriptscriptstyle T}}(s|\mathcal{H}_{1})$
(which depends on $\mathrm{SNR}$); such difference is explained since,
whichever $(\mathrm{SNR})_{\star}<+\infty$ is assumed, we have that
$\mathrm{SNR\rightarrow+\infty}$ in a large-system regime ($K\rightarrow+\infty$).
\item It can be verified that, when $(P_{D},P_{F})=(1,0)$, Eqs. (\ref{eq:MGFH1_kinf_IPC})
and (\ref{eq:MGFH0_kinf_IPC}) reduce to $\bar{\Phi}_{-\tilde{\Lambda}}^{{\scriptscriptstyle I}}(s|\mathcal{H}_{1})=(1-s)^{-N}$
and $\bar{\Phi}_{-\tilde{\Lambda}}^{{\scriptscriptstyle I}}(s|\mathcal{H}_{0})=(1+s)^{-N}$,
respectively. In this case $\tilde{\Lambda}^{I}|\mathcal{H}_{1}\sim\Gamma(N;1)$
and $-\tilde{\Lambda}^{I}|\mathcal{H}_{0}\sim\Gamma(N;1)$,\emph{
}thus\emph{ }leading to\emph{ ideal performance}%
\footnote{Note that $p_{\tilde{\Lambda}}^{I}(\lambda|\mathcal{H}_{1})\neq0$,
$\lambda\in\mathbb{R}^{+}$, and $p_{\tilde{\Lambda}}^{I}(\lambda|\mathcal{H}_{0})\neq0$,
$\lambda\in\mathbb{R}^{-}$, i.e. the supports of the two pdfs are
non-overlapping.%
}\emph{ }(i.e. the point $(P_{D_{0}},P_{F_{0}})=(1,0)$\emph{ }belongs
to the ROC)\emph{.} This is confirmed by observing that when $(P_{D,}P_{F})=(1,0)$
Eq. (\ref{eq:channel_model}) reduces to $\bm{y}=(\bm{H}\bm{1}_{K})\cdot\kappa+\bm{w}$,
$\kappa\in\{-1,1\}$; in this case $\Lambda{}_{{\scriptscriptstyle MRC}}$
represents the output of a standard MIMO-MRC combiner without beamforming
\cite{Love2003} and the limit $K\rightarrow+\infty$ under IPC determines
$\mathrm{SNR}\rightarrow+\infty$ at the combiner. This consideration
underlines the significant difference in terms of performance of the
MRC in the context of DF (where ($P_{D},P_{F})\neq(1,0)$) w.r.t.
its use in classic combining systems (where $(P_{D},P_{F})=(1,0)$).
\item The large system ROC in both IPC and TPC cases can be driven toward
the point $(P_{D_{0},}P_{F_{0}})=(1,0)$ by increasing also $N$.
Such a result agrees with Eq. (\ref{eq:large antenna array MRC}),
where on the other hand $K$ is kept finite and a large antenna array
($N\rightarrow+\infty$) is considered. In fact, in the latter case
the dependence on the specific channel realization vanishes, i.e.
both the IC and the CA cases coincide, and MRC performance approaches
those of a noisy counting rule, whose performance improves with $K$.
\end{itemize}

\subsection{Area under the ROC\label{sec:AUC_subsection}}

The AUC has been widely used as a global and synthetic measure of
performance in machine learning applications \cite{Bradley1997}.
Recently, it has been successfully applied to the performance analysis
of communication systems employing diversity techniques \cite{Atapattu2010,Atapattu2011}.
More specifically, in \cite{Atapattu2010} the AUC has been derived
in closed form and used for a theoretical comparison of several diversity
combining statistics in the classic communication framework. In the
case of non-coherent combining, an efficient formula for the AUC of
energy detection, based on the MGF and Cauchy's theorem, has been
recently proposed in \cite{Atapattu2011}. 

The AUC is defined on the basis of Eqs. (\ref{eq:CA_PF_0}) and (\ref{eq:CA_PD_0})
as follows:

\begin{align}
\mathrm{AUC} & \triangleq\intop_{0}^{1}P_{D_{0}}(\gamma)dP_{F_{0}}(\gamma)=-\intop_{-\infty}^{+\infty}P_{D_{0}}(\gamma)\frac{\partial P_{F_{0}}(\gamma)}{\partial\gamma}d\gamma.\label{eq:AUC_main_formula}
\end{align}
Note that, given the definition in Eq. (\ref{eq:AUC_main_formula}),
$\mathrm{AUC}\in[\frac{1}{2},1]$, i.e. the performances of any fusion
rule that exploits the measurements in a productive way cannot be
worse than\emph{ }a strategy based on \emph{flipping an unbiased coin}.
The information given by the $\mathrm{AUC}$ can be alternatively
expressed in terms of the \emph{Gini index,} denoted as $\mathrm{G}_{\mathrm{I}}$,
which is directly related to the $\mathrm{AUC}$ as follows \cite{Fawcett2004}:
\begin{equation}
\mathrm{G_{I}}\triangleq2\cdot\mathrm{AUC}-1\qquad\mathrm{G_{I}}\in[0,1].\label{eq:Gini_coeff_def}
\end{equation}
 Note that in our case the $\mathrm{AUC}$ depends on the local sensor
performance (i.e. $P(\ell(\bm{x})|\mathcal{H}_{i})$, $\mathcal{H}_{i}\in\mathcal{H}$),
the $\mathrm{SNR}$, the number of sensors $K$ and the number of
antennas $N$. Unfortunately, the evaluation of the $\mathrm{AUC}$
of a detection statistic, through MC simulations, suffers from high
computational complexity. In the following proposition we derive an
alternative expression for the $\mathrm{AUC}$ which allows its efficient
GC-based computation, similarly as shown for $P_{D_{0}}(\gamma)$
and $P_{F_{0}}(\gamma)$.
\begin{prop}
The AUC in Eq. (\ref{eq:AUC_main_formula}) can be expressed in the
alternative form\label{prop:MGF_AUC}
\begin{equation}
\mathrm{AUC}=\intop_{c-j\infty}^{c+j\infty}\frac{\Phi_{-\Lambda}(s|\mathcal{H}_{1})\Phi_{-\Lambda}(-s|\mathcal{H}_{0})}{2\pi j}\frac{ds}{s}.\label{eq:AUC_Laplace Domain}
\end{equation}
where $c$ is a positive constant in the RC of $\Phi_{-\Lambda}(s|\mathcal{H}_{1})$.\end{prop}
\begin{IEEEproof}
The proof is given in Appendix \ref{sec:Appendix_MGF_AUC}.
\end{IEEEproof}
Eq. (\ref{eq:AUC_main_formula}) is similar to the alternative expression
derived in \cite{Barrett1998}; however the difference here is that
Eq. (\ref{eq:AUC_main_formula}) is not formulated in terms of a contour
integral, which would require the evaluation of the residues of $\frac{\Phi_{-\Lambda}(s|\mathcal{H}_{1})\Phi_{-\Lambda}(-s|\mathcal{H}_{0})}{s}$,
through the \emph{Cauchy's theorem}. Differently, the AUC in Eq. (\ref{eq:AUC_main_formula})
can be computed exploiting the GC quadrature rules, analogously as
in Eqs. (\ref{eq:GC explicit formula_Pd0-Pf0}) and (\ref{eq:Complex Function quadrature Pd0-Pf0}),
through:
\begin{eqnarray}
\mathrm{AUC} & \approx & \frac{1}{\nu}\sum_{r=1}^{v/2}\left[\Re\{\psi(r)\}+\tau_{r}\Im\{\psi(r)\}\right];\label{eq:AUC_MGF_GCfinal}\\
\psi\left(r\right) & \triangleq & \Phi_{-\Lambda}(\mu_{r}|\mathcal{H}_{1})\Phi_{-\Lambda}(-\mu_{r}|\mathcal{H}_{0});\label{eq:AUC_MGF_GCfinal2}
\end{eqnarray}
where $\nu$, $\tau_{r}$ and $\mu_{r}$ have the same meaning as
in Eqs. (\ref{eq:GC explicit formula_Pd0-Pf0}) and (\ref{eq:Complex Function quadrature Pd0-Pf0}).

\section{System Design via Deflection Coefficients\label{sec:System Design - Def. Coefficients}}

We have shown in Section \ref{sec:MRC-Analytical-Resultsperformance}
that efficient computation of $P_{F_{0}}(\gamma)$ and $P_{D_{0}}(\gamma)$
can be obtained through Eqs. (\ref{eq:laplace_integral}), (\ref{eq:mgf_lambda_mrc})
and (\ref{eq:MGF_efficient_expansion}). Also, the AUC, which represents
a compact indicator for performance evaluation, can be evaluated through
Eqs. (\ref{eq:AUC_MGF_GCfinal}) and (\ref{eq:AUC_MGF_GCfinal2})
at low computational complexity. However, a compact and explicit metric
(i.e. independent on $\gamma$, as the $\mathrm{AUC}$, and available
in closed form), suitable for system design, would be desirable. The
deflection coefficient $D_{0}$ and its modified version $D_{1}$
are parameters%
\footnote{In the specific case of a mean-shifted Gauss-Gauss hypothesis testing,
they coincide and represent the $\mathrm{SNR}$ of the statistic under
Neyman-Pearson framework \cite{Kay1998}.%
} commonly employed in system design and analysis \cite{Picinbono1995},
while requiring only the first two order conditional moments. They
are defined as follows \cite{Chen2004,Picinbono1995,Quan2008}:
\begin{equation}
D_{i}\triangleq\frac{\left[\mathbb{E}\{\Lambda|\mathcal{H}_{1}\}-\mathbb{E}\{\Lambda|\mathcal{H}_{0}\}\right]^{2}}{\mathrm{var}\{\Lambda|\mathcal{H}_{i}\}},\quad i\in\{0,1\}.
\end{equation}
The expressions of $\mathbb{E}\{\Lambda_{{\scriptscriptstyle MRC}}|\mathcal{H}_{i}\}$
and $\mathrm{var}\{\Lambda_{{\scriptscriptstyle MRC}}|\mathcal{H}_{i}\}$
are given in closed form by the following proposition.%
\footnote{Hereinafter, we drop the dependence of $\ell(\cdot)$ w.r.t. $\bm{x}$
for ease of notation.%
} 
\begin{prop}
The mean and the variance of $\Lambda_{{\scriptscriptstyle MRC}}|\mathcal{H}_{i}$
are:\label{prop:DefCoeff}
\begin{align}
\mathbb{E}\{\Lambda_{{\scriptscriptstyle MRC}}|\mathcal{H}_{i}\} & =2\, N\,\mathbb{E}\{\ell|\mathcal{H}_{i}\}-KN;\\
\mathrm{var}\{\Lambda_{{\scriptscriptstyle MRC}}|\mathcal{H}_{i}\} & =K^{2}N(1+\frac{1}{2\,\mathrm{SNR}})-2\, KN\,\mathbb{E}\{\ell|\mathcal{H}_{i}\}\nonumber \\
 & +2N\,\mathbb{E}\{\ell^{2}|\mathcal{H}_{i}\}+4N^{2}\mathrm{var}\{\ell|\mathcal{H}_{i}\}.
\end{align}
\end{prop}
\begin{IEEEproof}
The moments of $\Lambda_{{\scriptscriptstyle MRC}}|\mathcal{H}_{i}$
are evaluated through the MGF definition \cite{Schwarz1966}: 
\begin{equation}
\mathbb{E}\{(\Lambda_{{\scriptscriptstyle MRC}})^{m}|\mathcal{H}_{i}\}=\left.\frac{d^{m}\left[\Phi_{-\Lambda_{MRC}}(s|\mathcal{H}_{i})\right]}{ds^{m}}\right|_{s=0}
\end{equation}
Hence, the first two order moments are obtained by setting $m=1$
and $m=2$, respectively. Finally, the variance is computed as $\mathrm{var}\{\Lambda_{{\scriptscriptstyle MRC}}|\mathcal{H}_{i}\}=\mathbb{E}\{\Lambda_{{\scriptscriptstyle MRC}}^{2}|\mathcal{H}_{i}\}-\mathbb{E}\{\Lambda_{{\scriptscriptstyle MRC}}|\mathcal{H}_{i}\}^{2}$.
\end{IEEEproof}
As a corollary we also report $D_{i}^{{\scriptscriptstyle MRC}}$
explicitly in the case of conditionally i.i.d. sensor decisions.
\begin{cor}
The deflection coefficients, when $P(\bm{x}|\mathcal{H}_{i})=\prod_{k=1}^{K}P(x_{k}|\mathcal{H}_{i})$,
$\mathcal{H}_{i}\in\mathcal{H}$, and $(P_{D,k},P_{F,k})=(P_{D},P_{F})$,
$k\in\mathcal{K},$ are given by:
\begin{align}
D_{i}^{{\scriptscriptstyle MRC}} & =\frac{4NK\left(P_{D}-P_{F}\right)^{2}}{K\,(1+\frac{1}{2\,\mathrm{SNR}})+2\left(2N+1-K\right)\rho_{i}},\quad i\in\{0,1\},\label{eq:D_i-iid}
\end{align}
where $\rho_{0}\triangleq P_{F}(1-P_{F})$ and $\rho_{1}\triangleq P_{D}(1-P_{D})$. 
\end{cor}
We will now analyze the qualitative behaviour of the deflection coefficients
(we will restrict our attention, for sake of simplicity, to $D_{i}^{{\scriptscriptstyle MRC}}$
in the conditionally i.i.d. case) in order to assess their efficacy
(in this specific problem) for analysis and design purposes. In fact
it is worth remarking that, as carefully specified in \cite{Picinbono1995},
an \emph{improvement in the deflection coefficients} in a generic
detection problem \emph{does not guarantee a corresponding improvement
in terms of} $\{P_{F_{0}},P_{D_{0}}\}$ and thus they should be used
with care. 

We start noticing that $D_{i}^{{\scriptscriptstyle MRC}}$ is strictly
increasing with $K$ under both IPC and TPC cases. Therefore, the
large-system (i.e. $K\rightarrow+\infty$) deflections
\begin{eqnarray}
\bar{D}_{i}^{{\scriptscriptstyle MRC},I} & = & \frac{4N(P_{D}-P_{F})^{2}}{1-2\rho_{i}},\label{eq:D_i_Kinf_IPC}\\
\bar{D}_{i}^{{\scriptscriptstyle MRC},T} & = & \frac{4N(P_{D}-P_{F})^{2}}{(1+\frac{1}{2\mathrm{SNR}})-2\rho_{i}},\quad i\in\{0,1\},\label{eq:D_i_Kinf_TPC}
\end{eqnarray}
represent the maximum attainable, when $N$ (and $\mathrm{SNR}$ under
TPC) is kept fixed. Some important observations are listed hereinafter:
\begin{itemize}
\item From inspection of Eqs. (\ref{eq:D_i_Kinf_IPC}) and (\ref{eq:D_i_Kinf_TPC}),
we have that $\bar{D}_{i}^{{\scriptscriptstyle MRC,T}}<\bar{D}_{i}^{{\scriptscriptstyle MRC,I}}<+\infty$,
i.e. the large-system deflection coefficients under both IPC and TPC
are finite, thus being in agreement with non-ideal performance shown
in Subsection \ref{sub:Large System WSN analysis}.
\item As $\mathrm{SNR}\rightarrow+\infty$ in Eq. (\ref{eq:D_i_Kinf_TPC}),
we have $\bar{D}_{i}^{{\scriptscriptstyle MRC},T}\rightarrow\bar{D}_{i}^{{\scriptscriptstyle MRC},I}$,
thus being in agreement with Fig. \ref{fig:Asymptotic-ROC-IPC TPC}
where CA-ROC under TPC approaches that under IPC as $\mathrm{SNR}$
increases.
\item Taking $N\rightarrow+\infty$ in Eqs. (\ref{eq:D_i_Kinf_IPC}) and
(\ref{eq:D_i_Kinf_TPC}) we get $\bar{D}_{i}^{{\scriptscriptstyle MRC},I}=\bar{D}_{i}^{{\scriptscriptstyle MRC},T}=+\infty$;
this result agrees with the ideal performance attainable in a large-system
regime when we let $N$ grow (cf. Fig. \ref{fig:Asymptotic-ROC-IPC TPC}).
\item If we set $(P_{D},P_{F})=(1,0)$ (i.e. the perfect sensor assumption)
in Eq. (\ref{eq:D_i_Kinf_IPC}) we obtain $\bar{D}_{i}^{{\scriptscriptstyle MRC},I}=4N$;
this\emph{ disagrees with the ideal performance }attained in a large-system
regime under IPC (see second bullet in\emph{ }Subsection \ref{sub:Large System WSN analysis}).
Such discrepancy is explained since, under the aforementioned assumptions,
we have $\tilde{\Lambda}^{I}|\mathcal{H}_{1}\overset{d}{\rightarrow}\Gamma(N;1)$
and $-\tilde{\Lambda}^{I}|\mathcal{H}_{0}\overset{d}{\rightarrow}\Gamma(N;1)$,
i.e. the pdfs $p_{\tilde{\Lambda}^{I}}(\lambda|\mathcal{H}_{0})$
and $p_{\tilde{\Lambda}^{I}}(\lambda|\mathcal{H}_{1})$ have \emph{a
non-zero variance but non-overlapping supports}%
\footnote{In fact, it can be easily verified that when $K\rightarrow+\infty$
and $(P_{D},P_{F})=(1,0)$, we have $E\{\tilde{\Lambda}^{I}|\mathcal{H}_{1}\}=N$,
$E\{\tilde{\Lambda}^{I}|\mathcal{H}_{0}\}=-N$ and $\mathrm{var}\{\tilde{\Lambda}^{I}|\mathcal{H}_{i}\}=N$. %
}\emph{.}
\end{itemize}
On the basis of the previous considerations it can be deduced that
$D_{i}^{{\scriptscriptstyle MRC}}$, $i\in\{0,1\}$, cannot be effectively
used for performance analysis of MRC, but that it can be rather suited
for system design, since it retains the same dependence on the WSN
parameters as the CA-ROC. For this reason we will use the (modified)
deflection as an optimization metric in order to obtain a choice of
the sensor threshold. We will formulate here the optimization w.r.t.
$P_{F}$, since we make the reasonable assumption that a \emph{one-to-one}
mapping between the local threshold and $P_{F}$ exists. More specifically,
exploiting Eq. (\ref{eq:D_i-iid}), we are interested in obtaining:
\begin{gather}
P_{F}^{*,i}\triangleq\arg\max_{P_{F}}D_{i}^{{\scriptscriptstyle MRC}}(P_{F})\\
=\arg\max_{P_{F}}\frac{4NK\cdot\left(P_{D}(P_{F})-P_{F}\right)^{2}}{K\,(1+\frac{1}{2\,\mathrm{SNR}})+2\cdot\left(2N+1-K\right)\cdot\rho_{i}(P_{F})}.\label{eq:optimization_Di}
\end{gather}
It can be noticed that $D_{i}^{{\scriptscriptstyle MRC}}(P_{F})$,
as it will be shown in Section \ref{sec:Simulations-Results} through
simulations, is \emph{quasi-concave }(i.e. unimodal) \cite{Boyd2004}.\emph{
}Thus local-optimization procedures, based on standard \emph{quasi-convex
programming}, can be easily devised in order to obtain\emph{ $P_{F}^{*,i}$.}
However, the derivation and comparison of such procedures is outside
the scope of this work.\emph{ }The improvement in terms of performance
on the CA-ROC, attained with such optimization, will be verified in
Section \ref{sec:Simulations-Results}.

\section{Numerical Results\label{sec:Simulations-Results}}

In this section we verify and analyze the theoretical results obtained
in Sections \ref{sec:MRC-Analytical-Resultsperformance} and \ref{sec:System Design - Def. Coefficients}.
For simplicity and w.l.o.g. we consider conditionally i.i.d. sensor
decisions, i.e. $P(\bm{x}|\mathcal{H}_{i})=\prod_{k=1}^{K}P(x_{k}|\mathcal{H}_{i})$,
$\mathcal{H}_{i}\in\mathcal{H}$, and $(P_{D,k},P_{F,k})=(P_{D},P_{F})$,
$k\in\mathcal{K}$. Unless differently stated, we assume $(P_{D},P_{F})\triangleq(0.5,0.05)$,
as adopted in \cite{Chen2004,Lei2010} for fusion rules comparison
over PAC.
\begin{figure}
\begin{centering}
\subfloat[$K=50$, $N=1$.]{\centering{}\includegraphics[width=0.5\columnwidth]{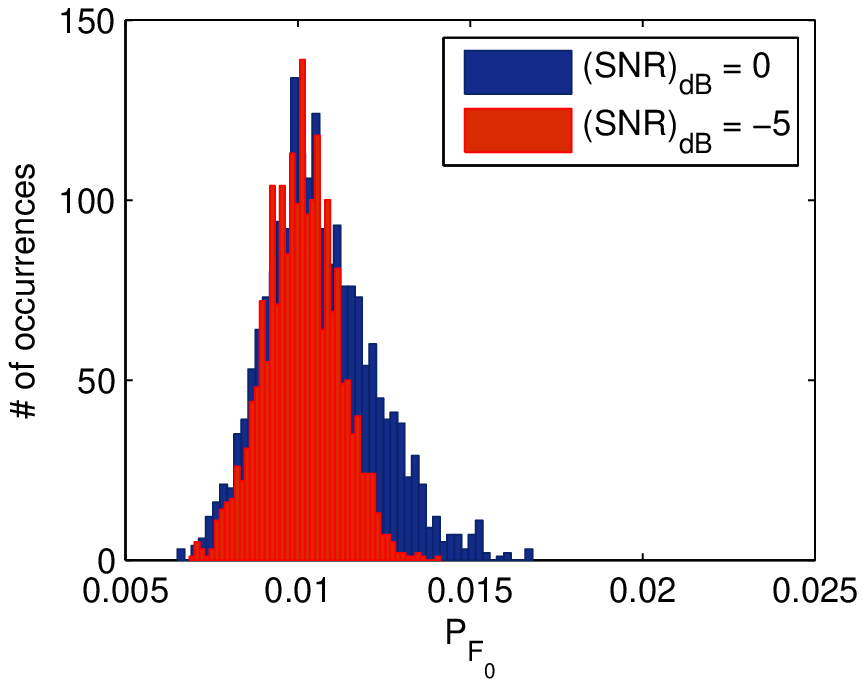}}\subfloat[$K=50$, $N=2$.]{\centering{}\includegraphics[width=0.5\columnwidth]{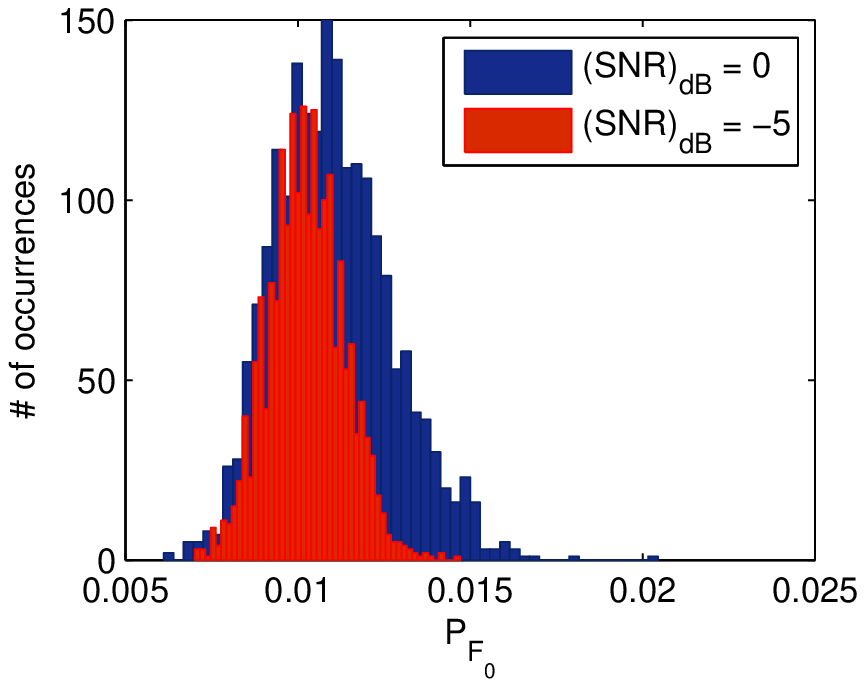}}
\par\end{centering}

\begin{centering}
\subfloat[$K=100$, $N=1$.]{\centering{}\includegraphics[width=0.5\columnwidth]{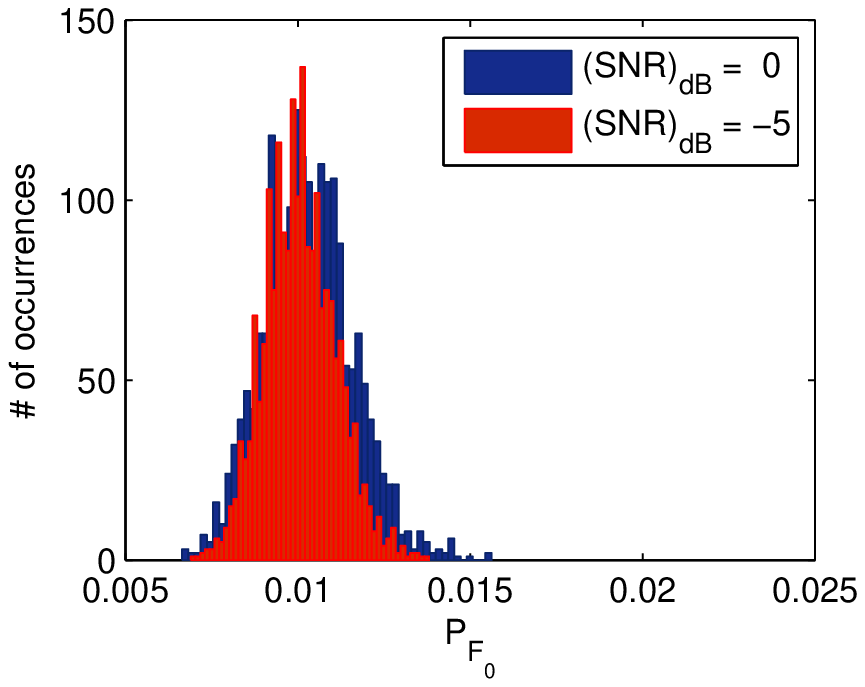}}\subfloat[$K=100$, $N=2$.]{\centering{}\includegraphics[width=0.5\columnwidth]{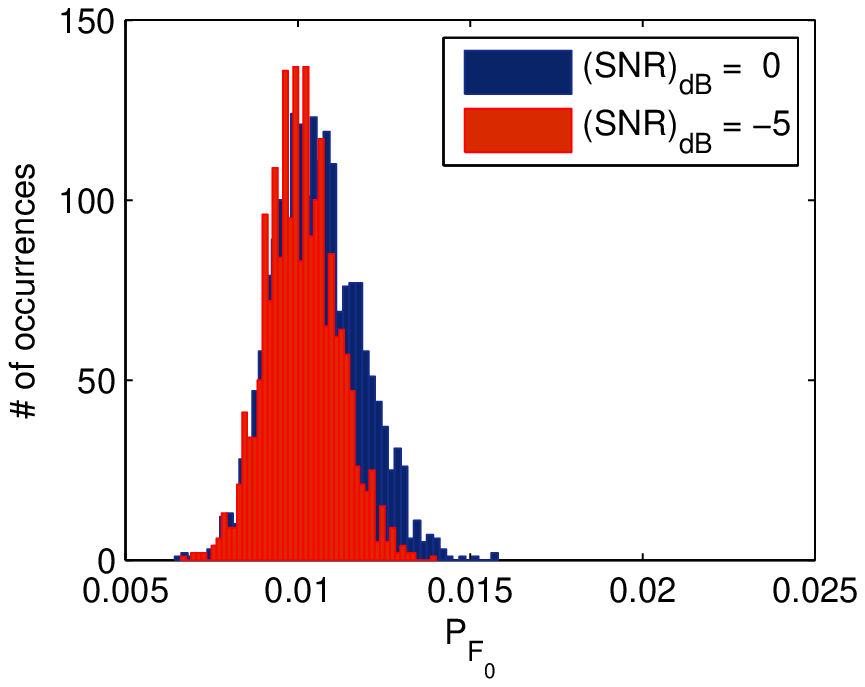}}
\par\end{centering}

\caption{Threshold choice $\breve{\gamma}$ for target $\breve{P}_{F_{0}}=0.01$;
$(P_{D,k},P_{F,k})=(0.5,0.05)$, $k\in\mathcal{K}$.\label{fig:Threshold Choice histograms}}
\end{figure}

\emph{IC false-alarm rate with threshold }$\breve{\gamma}$: In Fig.
\ref{fig:Threshold Choice histograms} we assess the accuracy of $\breve{\gamma}$
(given by Eq. (\ref{eq:low-SNR large-system threshold})) through
MC simulations; we choose here $\breve{P}_{F_{0}}=0.01$. The experiment
is conducted as follows. We generate $(2\cdot10^{3})$ realizations
of $\bm{H}$ and, for each of these, we generate $10^{4}$ realizations
of $(\bm{w},\bm{x})$ to obtain an estimate of $P_{F_{0}}(\breve{\gamma},\bm{H})$.
Finally an histogram of the r.v. $P_{F_{0}}(\breve{\gamma},\bm{H})$
is obtained by considering all the realizations of $\bm{H}$. We report
the cases corresponding to $N\in\{1,2\}$, $(\mathrm{SNR})_{dB}\in\{-5,0\}$
and $K\in\{50,100\}$. Since $\breve{\gamma}$ is a low-$\mathrm{SNR}$
and large-system approximation, as the $\mathrm{SNR}$ decreases and
$K$ increases the histogram approaches a delta function centered
at the desired IC false-alarm rate; furthermore it can be seen that
a low-SNR assures unbiasedness of the estimate, while increasing $K$
reduces the variance of the histogram. 

\begin{figure}
\begin{centering}
\includegraphics[width=0.99\columnwidth]{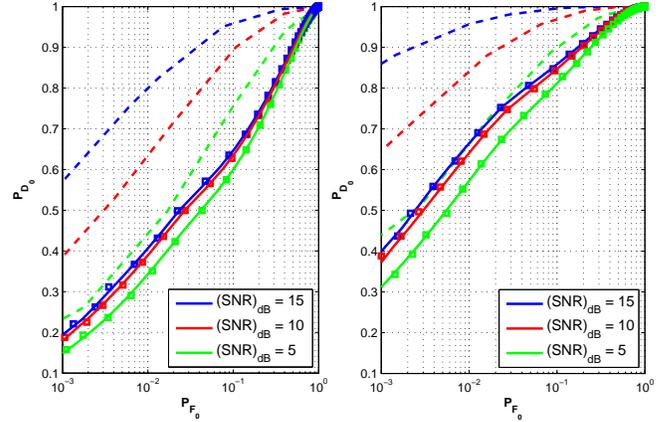}
\par\end{centering}

\caption{Max-Log (dashed lines), GC-based MRC (solid lines) and MC-based MRC
($\square$ markers)  CA-ROC evaluation. WSN with $K=8$, $N=2$,
$(\mathrm{SNR}){}_{dB}\in\{5,10,15\}$. $(P_{D,k},P_{F,k})=(0.5,0.05)$
for scenario (a) (left plot), $(P_{D,k},P_{F,k})=(0.7,0.05)$ for
scenario (b) (right plot), $k\in\mathcal{K}$.\label{fig:ROC_an_performance}}
\end{figure}

\emph{MC vs GC rules CA-ROC}: Fig. \ref{fig:ROC_an_performance} shows
the CA-ROC of the MRC rule in a WSN with $K=8$ and $N=2$ at the
DFC in two scenarios with different $P_{D}$ (we fix $P_{F}=0.05$):
(a) $P_{D}=0.5$; (b) $P_{D}=0.7$. For sake of completeness we also
report the CA-ROC of Max-Log fusion rule (in dashed lines), which
represents an approximated and efficient implementation of the optimum
in Eq. (\ref{eq:optimum_llr}), but exhibiting negligible performance
loss \cite{Ciuonzo2012}. For each scenario we report the performance
at $(\mathrm{SNR})_{dB}\in\{5,10,15\}$. Solid lines represent GC-based
computation of MRC CA-ROC ($\nu=10^{3}$ for each value of $\gamma$),
while square markers represent the corresponding MC-based evaluation
($10^{5}$ runs for each value of $\gamma$), in the two scenarios
respectively. It is apparent how the proposed approach perfectly matches
the MC simulations, while requiring dramatically reduced computational
resources (the complexity is in fact reduced \emph{roughly by two
orders of magnitude}%
\footnote{Even if the two approaches are not directly comparable, we observe
that in the former case the complexity is proportional to $\nu$,
while with MC-based computation it is proportional to the number of
runs. Furthermore, when using the GC-based computation, $\nu$ could
be further reduced through an optimized choice of $c$.%
}). Finally, it is also apparent the increasing performance loss of
MRC with respect to Max-Log as the $\mathrm{SNR}$ increases (since
MRC is a low-$\mathrm{SNR}$ approximation of the optimum rule). 

\begin{figure}[t]
\centering{}\includegraphics[width=0.95\columnwidth]{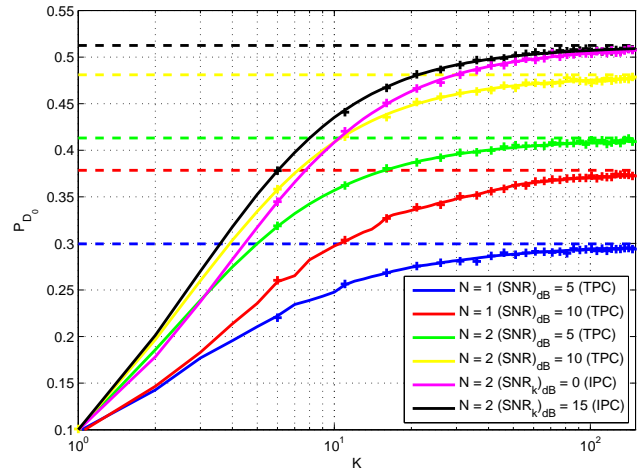}\caption{$P_{D_{0}}$ vs $K$, $P_{F_{0}}=0.01$; $(P_{D,k},P_{F,k})=(0.5,0.05)$,
$k\in\mathcal{K}$. Solid lines denote GC-based $P_{D_{0}}$, while
plus ($+$) markers refer to MC-based $P_{D_{0}}$; finally dotted
lines denote GC-based $P_{D_{0}}$ evaluation through large-system
conditional MGFs. \label{fig: P_D0 vs K}}
\end{figure}

\emph{$P_{D_{0}}$ vs }$K$: In Fig. \ref{fig: P_D0 vs K} we report
$P_{D_{0}}$ as a function of the number of sensors $K$, under $P_{F_{0}}=0.01$;
we show the performance for several configurations of $(\mathrm{SNR})_{dB}$
under TPC (resp. $(\mathrm{SNR}_{\star})_{dB}$ under IPC) and $N$.
Firstly, the figure confirms that GC-based computation ($\nu=10^{3}$
for each value of $\gamma$ of each considered CA-ROC, plotted with
solid lines) perfectly matches the MC simulations ($10^{5}$ runs
for each value of $\gamma$ of each considered CA-ROC, plotted with
plus markers). Secondly, as $K$ increases, there is a saturation
effect in $P_{D_{0}}$, which converges to a value smaller than $1$.
Such a result is perfectly predicted through GC-based computation
($\nu=10^{3}$ for each value of $\gamma$ of each considered CA-ROC,
plotted in dotted lines) of large-system conditional MGFs in Eqs.
(\ref{eq:MGFH1_kinf_IPC}-\ref{eq:MGFH0_kinf_TPC}), thus confirming
Proposition \ref{prop:largesystemMGFs}. Finally, we remark that a
similar behaviour has been observed when considering the overall CA-ROC
performance, expressed in terms of $\mathrm{G_{I}}$.

\emph{$\mathrm{G}_{\mathrm{I}}$} \emph{vs} $(K,N)$: In Fig. \ref{fig:AUC-vs-(K,N)}
we illustrate $\mathrm{G_{I}}$ as a function of both $K$ and $N$
to investigate how performance saturation can be avoided. We consider
both IPC ($(\mathrm{SNR}_{\star})_{dB}=10$) and TPC ($\mathrm{(SNR})_{dB}=10$),
in Figs. \ref{fig: AUC vs (K,N) - IPC} and \ref{fig: AUC vs (K,N) - TPC},
respectively; it is worth remarking that similar behaviours have been
observed for different values of $(\mathrm{SNR})_{dB}$ and $(\mathrm{SNR}_{\star})_{dB}$.
Each surface is obtained exploiting GC-based computation (only $\nu=64$)
in Eqs. (\ref{eq:AUC_MGF_GCfinal}) and (\ref{eq:AUC_MGF_GCfinal2});
the corresponding MC-based $\mathrm{G_{I}}$ is not reported for sake
of clarity, since a \emph{perfect match has been noticed}. The surfaces
show that when either $K$ or $N$ is kept fixed and the other parameter
grows we cannot attain $\mathrm{\mathrm{G_{I}}=1}$ (i.e. ideal performance)
both in IPC and TPC cases. Differently, we observe that ideal performances
are achievable when both the parameters increase simultaneously. For
this reason, in Fig. \ref{fig:Gc vs (K,alpha*K)} we analyze $\mathrm{G_{I}}$
after fixing $N=\alpha K$ (we consider $\alpha\in\{\frac{1}{2},\frac{1}{4},\frac{1}{8}\}$,
since it is reasonable to assume that typically $K>N$) and let $K$
grow. It is apparent that in this setup: ($i$) we can achieve ideal
performance as $K$ increases; $(ii)$ the value of $K$ needed to
achieve ideal performance\emph{ }decreases as $\alpha\rightarrow1$;
($iii$) the performance of IPC and TPC cases are roughly the same
as $K$ grows, since increasing $N$ corresponds to an increase of
the received SNR, \emph{independently} on the specific power constraint
assumed. 
\begin{figure*}
\subfloat[ $(\mathrm{SNR}_{\star})_{dB}=10$ (IPC case).\label{fig: AUC vs (K,N) - IPC}]{\includegraphics[width=0.31\paperwidth]{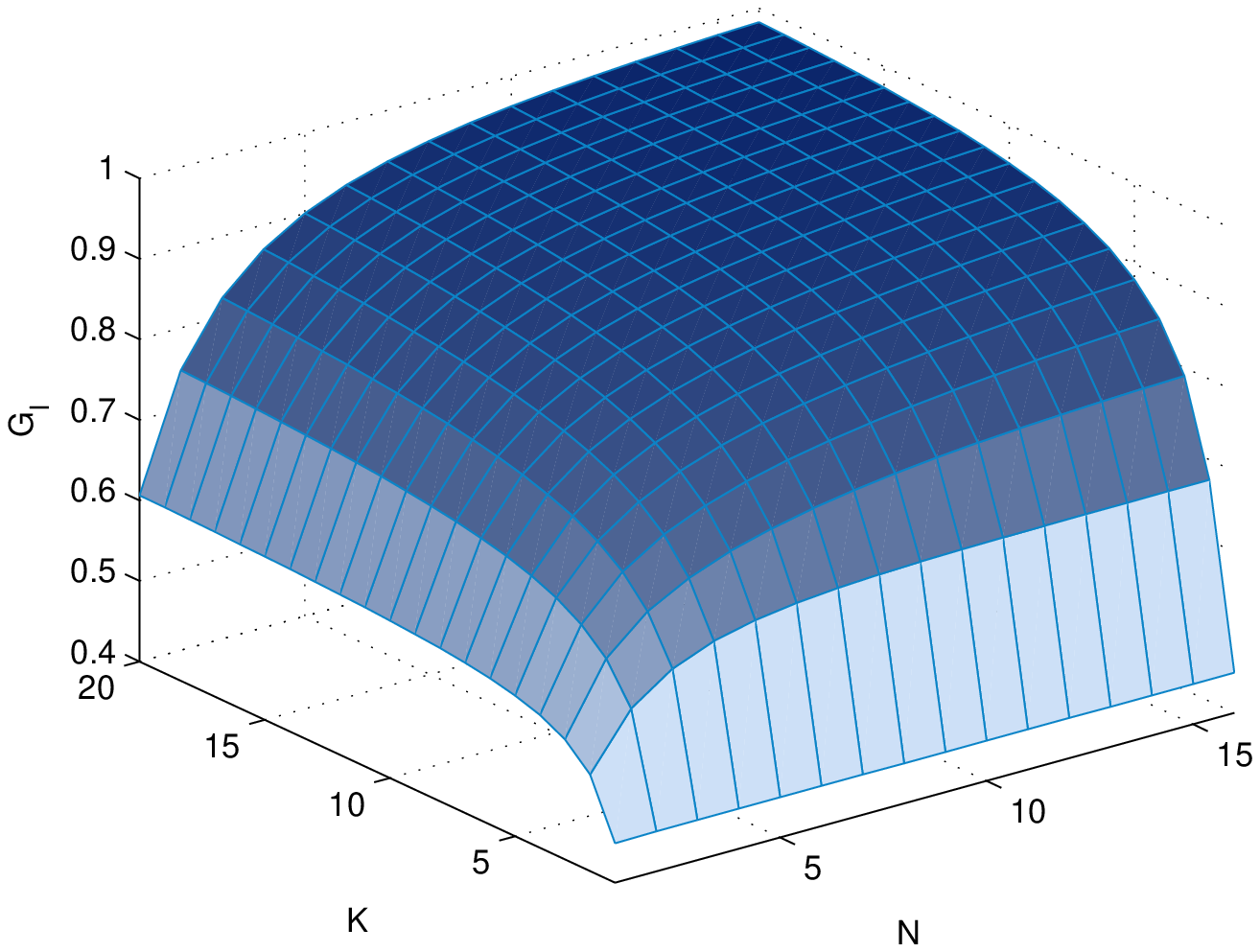}

}\subfloat[$(\mathrm{SNR})_{dB}=10$ (TPC case).\label{fig: AUC vs (K,N) - TPC}]{\includegraphics[width=0.31\paperwidth]{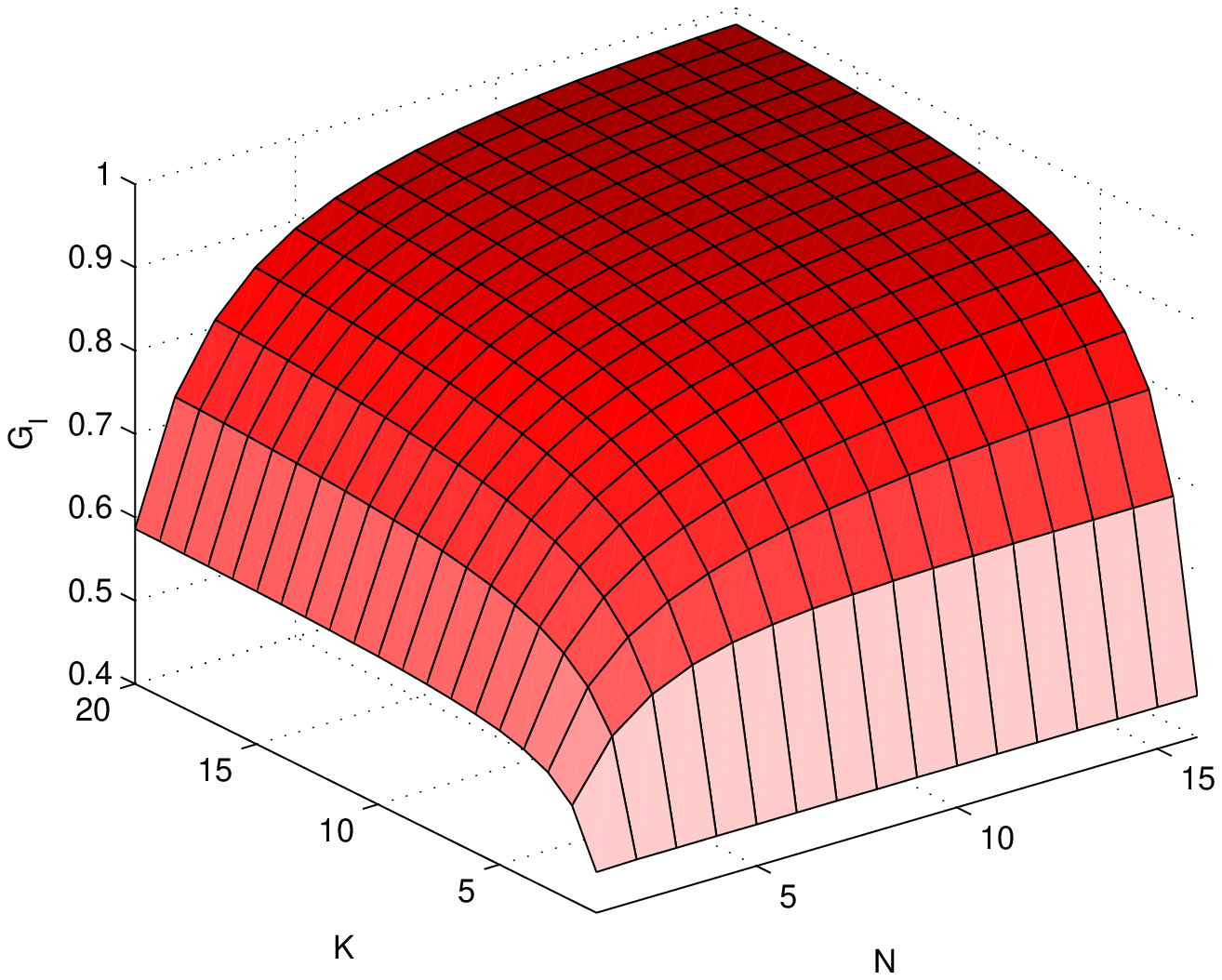}

}\subfloat[$N=\alpha K$, $(\mathrm{SNR}_{\star})_{dB}=10$ (resp. $(\mathrm{SNR})_{dB}=10$)
for IPC (resp. TPC) case.\label{fig:Gc vs (K,alpha*K)}]{\includegraphics[width=0.19\paperwidth]{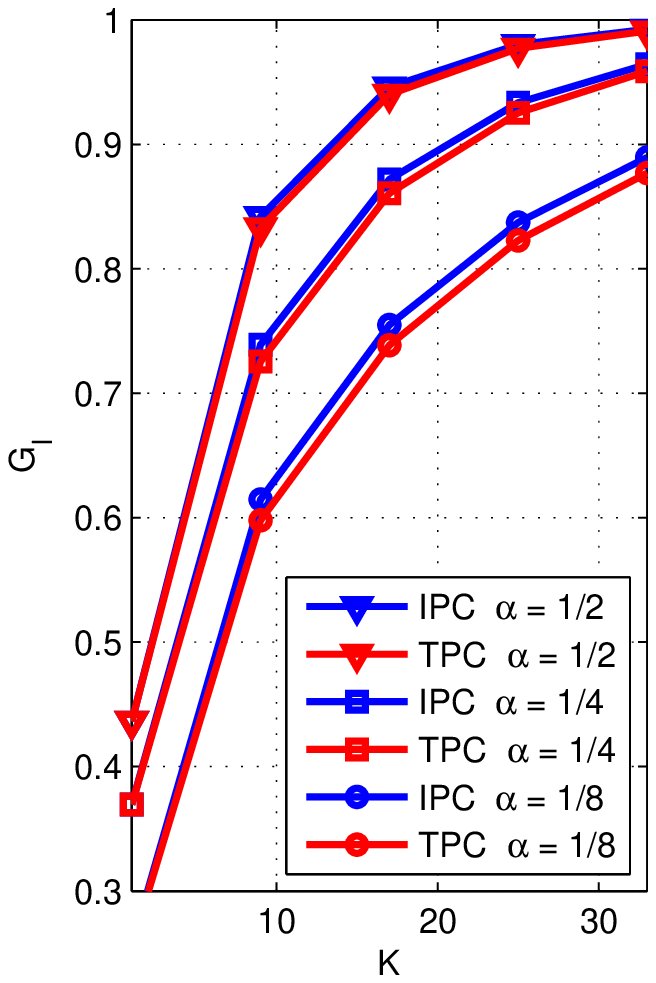}

}\caption{$\mathrm{G_{I}}$ vs $(K,N)$; $(P_{D,k},P_{F,k})=(0.5,0.05)$, $k\in\mathcal{K}$.\label{fig:AUC-vs-(K,N)}}
\end{figure*}

\emph{Sensor threshold optimization via $D_{i}^{{\scriptscriptstyle MRC}}$}:
In Fig. \ref{fig: Pd0 vs N - def opt - SNR =00003D 5 dB} we illustrate
the effect of a $D_{i}^{{\scriptscriptstyle MRC}}$-optimized choice
for $P_{F}$. We consider the scenarios $K\in\{15,50\}$ and $(\mathrm{SNR})_{dB}=5$
under TPC (resp. $(\mathrm{SNR}_{\star})_{dB}=5$ under IPC). We assume
that the generic sensor is a detector of a \emph{change-in-variance}
\cite{Kay1998}, whose closed-form expression is given by $P_{D}(P_{F})=P_{F}^{\nicefrac{1}{(1+\mathrm{SNR}_{obs})}}$,
where $\mathrm{SNR}_{obs}$ denotes \emph{sensing} SNR; we choose
$(\mathrm{SNR}_{obs})_{dB}=5$. We compare the performance when: ($i$)
$P_{F}=0.05$ (non-optimized choice); $(ii)$ $P_{F}=P_{F}^{*,1}$
(i.e. $D_{1}^{{\scriptscriptstyle MRC}}$-optimized, recall Eq. (\ref{eq:optimization_Di}));
($iii)$ $P_{F}=P_{F}^{*,0}$ (i.e. $D_{0}^{{\scriptscriptstyle MRC}}$-optimized,
recall Eq. (\ref{eq:optimization_Di})). In Figs. \ref{fig:Deflections IPC case}
and \ref{fig:Deflections TPC case} we show the corresponding $D_{i}^{{\scriptscriptstyle MRC}}(P_{F})$
under IPC and TPC, respectively, for all the scenarios considered
($D_{0}^{{\scriptscriptstyle MRC}}$ and $D_{1}^{{\scriptscriptstyle MRC}}$
in solid and dotted lines, respectively; $K=15$ and $K=50$ with
plus and circle markers, respectively); all the curves underline \emph{quasi-concavity}
of $D_{i}^{{\scriptscriptstyle MRC}}(P_{F})$. In Figs. \ref{fig:Pd0 vs N - def. optimization - IPC}
and \ref{fig:Pd0 vs N - def. optimization - TPC} we show $P_{D_{0}}$
(GC-based computation, $\nu=10^{3}$ for each value of $\gamma$ of
each considered CA-ROC) as a function of the number of antennas $N$,
under $P_{F_{0}}=0.01$, in IPC and TPC case, respectively. First,
it is apparent that deflection-based optimization of $P_{F}$ becomes
effective, in comparison to a non-optimized choice, as $N$ grows.
Furthermore, the improvement is more pronounced in the case $K=50$
where the choice becomes effective for small $N$ (i.e. $N>2$ when
$P_{F}=P_{F}^{*,1}$); therefore such threshold optimization is \emph{best-suited}
for a large-system. For example, when choosing $P_{F}=P_{F}^{*,1}$,
a $10\%$ improvement of $P_{D_{0}}$ is achieved in a configuration
with $(K,N)=(50,4)$. Also, it is observed that the choice $P_{F}=P_{F}^{*,1}$
is more convenient w.r.t. $P_{F}=P_{F}^{*,0}$ for this setup; this
is due to the higher detection sensitivity ensured when optimizing
$P_{F}$ w.r.t. $D_{1}^{{\scriptscriptstyle MRC}}$, in a Neyman-Pearson
scenario (i.e. a fixed $P_{F_{0}}$). Finally, it is worth noticing
that performance improvement effect is similar under both IPC and
TPC.

\begin{center}
\begin{figure*}[t]
\begin{centering}
\subfloat[$D_{i}^{{\scriptscriptstyle MRC}}$ vs $P_{F}$; IPC case, $(\mathrm{SNR}_{\star})_{dB}=5$.
Plus ($+$) and circle ($\circ$) markers refer to $K=15$ and $K=50$,
respectively.\label{fig:Deflections IPC case}]{\includegraphics[width=0.34\paperwidth]{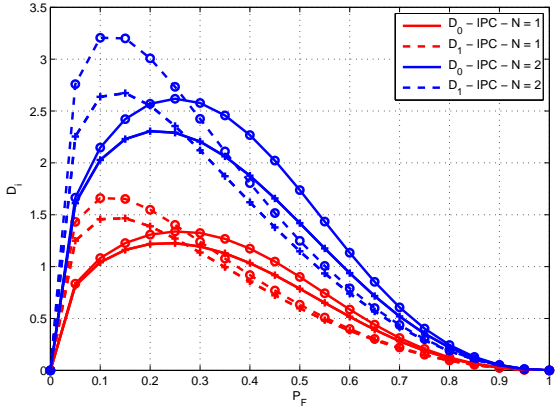}}\quad{}\subfloat[$P_{D_{0}}$ vs $N$; $P_{F_{0}}=0.01$. WSN with $K\in\{15,50\}$;
IPC case, $(\mathrm{SNR}_{\star})_{dB}=5$. Square ($+$) markers
refer to $P_{F}=0.05$ (non-optimized), diamonds $(\diamond)$ to
$P_{F}=P_{F}^{*,0}$ and cross markers ($\times$) to $P_{F}=P_{F}^{*,1}$.
\label{fig:Pd0 vs N - def. optimization - IPC}]{\includegraphics[width=0.35\paperwidth]{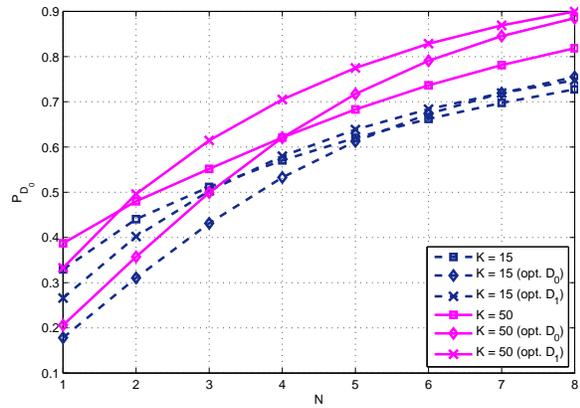}

}
\par\end{centering}

\begin{centering}
\subfloat[$D_{i}^{{\scriptscriptstyle MRC}}$ vs $P_{F}$; TPC case, $(\mathrm{SNR})_{dB}=5$.
Plus ($+$) and circle ($\circ$) markers refer to $K=15$ and $K=50$,
respectively.\label{fig:Deflections TPC case} ]{\includegraphics[width=0.34\paperwidth]{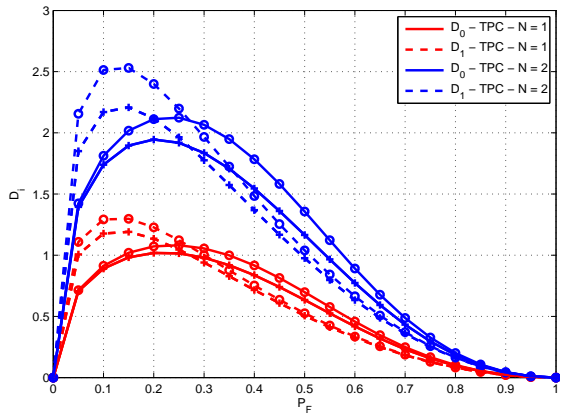}}\quad{}\subfloat[$P_{D_{0}}$ vs $N$; $P_{F_{0}}=0.01$. WSN with $K\in\{15,50\}$;
TPC case, $(\mathrm{SNR})_{dB}=5$. Square ($+$) markers refer to
$P_{F}=0.05$ (non-optimized), diamonds $(\diamond)$ to $P_{F}=P_{F}^{*,0}$
and cross markers ($\times$) to $P_{F}=P_{F}^{*,1}$. \label{fig:Pd0 vs N - def. optimization - TPC}]{\includegraphics[width=0.35\paperwidth]{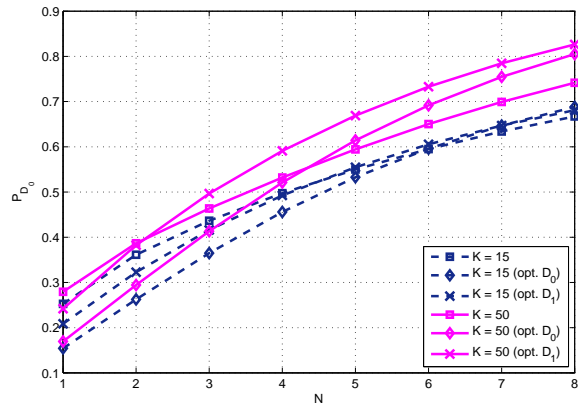}

}
\par\end{centering}

\caption{Performance comparison of $D_{i}^{{\scriptscriptstyle MRC}}$-optimized
vs non-optimized $P_{F}$ for MRC. $(\mathrm{SNR}_{obs})_{dB}=5$.\label{fig: Pd0 vs N - def opt - SNR =00003D 5 dB}}
\end{figure*}

\par\end{center}

\section{Conclusions\label{sec:Conclusions}}

In this paper we presented a performance analysis of MRC DF rule over
MIMO channels. We derived an approximate threshold choice with reduced
requirements on system knowledge, useful at low-SNR and in a large-system
regime, exploiting IC probabilities of false alarm and detection.
Also, the CA conditional MGF was derived in closed form in order to
compute efficiently the CA system probabilities of false alarm and
detection via GC rules. The explicit expression of CA conditional
MGFs was also exploited to demonstrate that ideal performance is not
attained in large-system limit under both IPC and TPC. The CA conditional
MGF was also used to derive an efficient computation of the AUC of
the proposed statistic, which was shown to be suited for synthetic
performance analysis w.r.t the WSN parameters. The AUC analysis also
showed how performance saturation can be avoided by increasing accordingly
the number of sensors and antennas. Finally, the (modified) deflection
coefficient was derived in closed form and it was shown to be effective
as an optimization metric for the local threshold choice, when the
DFC has multiple antennas. Such effect was shown to be more pronounced
when the number of sensors is large.

\section{Acknowledgements}

The authors would like to express their sincere gratitude to the Associate
Editor and the anonymous reviewers for taking their time into reviewing
this manuscript and providing comments that contributed to improve
the quality and the readability of the manuscript.

\appendices{}

\section{Proof of Proposition \ref{prop:threshold low-SNR}\label{sec:Appendix_Proposition-threshold-lowsnr}}

In the first part of this Appendix we will derive the closed form
expression for $p_{\Lambda_{MRC}}(\lambda|\bm{H},\mathcal{H}_{i})$,
$\lambda\in\mathbb{R}$, $\mathcal{H}_{i}\in\mathcal{H}$. From inspection
of Eqs. (\ref{eq:optimum_llr}) and (\ref{eq:MRC}) it can be shown
that, since the pdf of $\bm{y}|\bm{H},\mathcal{H}_{i}$ is a Gaussian
mixture with weights equal to $P(\bm{x}|\mathcal{H}_{i})$, also $\Lambda_{{\scriptscriptstyle MRC}}|\bm{H},\mathcal{H}_{i}$
will be distributed according to a Gaussian mixture with the same
weights (as the MRC rule represents a widely-linear transformation
of $\bm{y}$ \cite{Schreier2010}), that is
\begin{gather}
\Lambda_{{\scriptscriptstyle MRC}}|\bm{H},\mathcal{H}_{i}\sim\label{eq:IC_MRC_PDF}\\
\sum_{\bm{x}\in\mathcal{X}^{K}}P(\bm{x}|\mathcal{H}_{i})\mathcal{N}(E\{\Lambda_{{\scriptscriptstyle MRC}}|\bm{H},\bm{x}\},\mathrm{var}\{\Lambda_{{\scriptscriptstyle MRC}}|\bm{H},\bm{x}\}).\nonumber 
\end{gather}
To obtain a complete characterization of Eq. (\ref{eq:IC_MRC_PDF})
we now evaluate the mean and the variance of each component of the
mixture; for this purpose let us define $\tilde{\bm{z}}_{{\scriptscriptstyle MRC}}\triangleq$
$\frac{1}{2}\left[\begin{array}{cc}
\bm{z}_{{\scriptscriptstyle MRC}}^{t} & \bm{z}_{{\scriptscriptstyle MRC}}^{\dagger}\end{array}\right]^{t}$, $\tilde{\bm{H}}\triangleq\left[\begin{array}{cc}
\bm{H}^{t} & \bm{H}^{\dagger}\end{array}\right]^{t}$ , $\tilde{\bm{y}}\triangleq\left[\begin{array}{cc}
\bm{y}^{t} & \bm{y}^{\dagger}\end{array}\right]^{t}$ and $\tilde{\bm{w}}\triangleq\left[\begin{array}{cc}
\bm{w}^{t} & \bm{w}^{\dagger}\end{array}\right]^{t}$. The mean of $\Lambda_{{\scriptscriptstyle MRC}}|\bm{H},\bm{x}$
is obtained as follows
\begin{eqnarray}
\mathbb{E}\{\Lambda_{{\scriptscriptstyle MRC}}|\bm{H},\bm{x}\} & = & \mathbb{E}\{\tilde{\bm{z}}_{{\scriptscriptstyle MRC}}^{\dagger}\tilde{\bm{y}}|\bm{H},\bm{x}\}\\
 & = & \tilde{\bm{z}}_{{\scriptscriptstyle MRC}}^{\dagger}\tilde{\bm{H}}\bm{x}\\
 & = & \Re(\bm{z}_{{\scriptscriptstyle MRC}}^{\dagger}\bm{H}\bm{x}),\label{eq:meanistMRC}
\end{eqnarray}
where we exploited $\tilde{\bm{y}}=\tilde{\bm{H}}\bm{x}+\tilde{\bm{w}}$
and $\mathbb{E}\{\tilde{\bm{w}}\}=\bm{0}_{2N}$. Differently, the
variance is evaluated as
\begin{align}
\mathrm{var}\{\Lambda_{{\scriptscriptstyle MRC}}|\bm{H},\bm{x}\} & =\mathbb{E}\left\{ \left\Vert \tilde{\bm{z}}_{{\scriptscriptstyle MRC}}^{\dagger}\tilde{\bm{y}}-\tilde{\bm{z}}_{{\scriptscriptstyle MRC}}^{\dagger}\tilde{\bm{H}}\bm{x}\right\Vert ^{2}|\bm{H},\bm{x}\right\} \\
 & =\mathbb{E}\left\{ \left\Vert \tilde{\bm{z}}_{{\scriptscriptstyle MRC}}^{\dagger}\tilde{\bm{w}}\right\Vert ^{2}|\bm{H}\right\} \\
 & =\tilde{\bm{z}}_{{\scriptscriptstyle MRC}}^{\dagger}\mathbb{E}\left\{ \tilde{\bm{w}}\tilde{\bm{w}}^{\dagger}\right\} \tilde{\bm{z}}_{{\scriptscriptstyle MRC}}\\
 & =\frac{\sigma_{w}^{2}}{2}\left\Vert \bm{z}_{{\scriptscriptstyle MRC}}\right\Vert ^{2}\label{eq:varistMRC}
\end{align}
since $\bm{w}$ is independent on both $\bm{H}$ and $\bm{x}$ and
$\mathbb{E}\{\tilde{\bm{w}}\tilde{\bm{w}}^{\dagger}\}=\sigma_{w}^{2}\bm{I}_{2N}$.
Direct substitution of Eqs. (\ref{eq:meanistMRC}) and (\ref{eq:varistMRC})
in Eq. (\ref{eq:IC_MRC_PDF}) provides Eq.~(\ref{eq:istc_pdf_MRC}).

We will now prove Proposition \ref{prop:threshold low-SNR}. We start
noticing that at low-$\mathrm{SNR}$ the components of the mixture
will be concentrated and thus we can rely on the \emph{Gaussian moment
matching }\cite{Bar-Shalom2002} to approximate the pdf in Eq. (\ref{eq:IC_MRC_PDF})
as
\begin{align}
\Lambda_{{\scriptscriptstyle MRC}}|\bm{H},\mathcal{H}_{i} & \overset{{\scriptscriptstyle approx.}}{\sim}\mathcal{N}\left(\mathbb{E}\{\Lambda_{{\scriptscriptstyle MRC}}|\bm{H},\mathcal{H}_{i}\},\mathrm{var}\{\Lambda_{{\scriptscriptstyle MRC}}|\bm{H},\mathcal{H}_{i}\}\right).\label{eq:GMM_MRC}
\end{align}
To accomplish this task we need to evaluate the mean and variance
of $\Lambda_{{\scriptscriptstyle MRC}}|\bm{H},\mathcal{H}_{i}$. The
mean is obtained as\emph{ 
\begin{align}
\mathbb{E}\{\Lambda_{{\scriptscriptstyle MRC}}|\bm{H},\mathcal{H}_{i}\} & =\sum_{\bm{x}\in\mathcal{X}^{K}}\mathbb{E}\{\Lambda_{{\scriptscriptstyle MRC}}|\bm{H},\bm{x}\}P(\bm{x}|\mathcal{H}_{i})\label{eq:approx_ic_PF0}\\
 & =\sum_{\bm{x}\in\mathcal{X}^{K}}\Re(\bm{z}_{{\scriptscriptstyle MRC}}^{\dagger}\bm{H}\bm{x})P(\bm{x}|\mathcal{H}_{i})\\
 & =\Re\left(\bm{z}_{{\scriptscriptstyle MRC}}^{\dagger}\bm{H}\mathbb{E}\{\bm{x}|\mathcal{H}_{i}\}\right)
\end{align}
}where we exploited the explicit expression of $E\{\Lambda_{{\scriptscriptstyle MRC}}|\bm{H},\bm{x}\}$
in Eq. (\ref{eq:meanistMRC}). Differently, the variance is obtained
as \emph{
\begin{gather}
\mathrm{var}\{\Lambda_{{\scriptscriptstyle MRC}}|\bm{H},\mathcal{H}_{i}\}=\nonumber \\
\sum_{\bm{x}\in\mathcal{X}^{K}}\mathbb{E}\{\left\Vert \Lambda_{{\scriptscriptstyle MRC}}-\tilde{\bm{z}}_{{\scriptscriptstyle MRC}}^{\dagger}\tilde{\bm{H}}\mathbb{E}\{\bm{x}|\mathcal{H}_{i}\}\right\Vert ^{2}|\bm{H},\bm{x}\}P(\bm{x}|\mathcal{H}_{i})=\\
\sum_{\bm{x}\in\mathcal{X}^{K}}\mathbb{E}\{\left\Vert \tilde{\bm{z}}_{{\scriptscriptstyle MRC}}^{\dagger}\cdot\left(\tilde{\bm{y}}-\tilde{\bm{H}}\mathbb{E}\{\bm{x}|\mathcal{H}_{i}\}\right)\right\Vert ^{2}|\bm{H},\bm{x}\}P(\bm{x}|\mathcal{H}_{i})=
\end{gather}
\begin{gather}
\sum_{\bm{x}\in\mathcal{X}^{K}}(\mathbb{E}\{\left\Vert \tilde{\bm{z}}_{{\scriptscriptstyle MRC}}^{\dagger}\tilde{\bm{H}}\left(\bm{x}-\mathbb{E}\{\bm{x}|\mathcal{H}_{i}\}\right)\right\Vert ^{2}|\bm{H},\bm{x}\}+\nonumber \\
\mathbb{E}\{\left\Vert \tilde{\bm{z}}_{{\scriptscriptstyle MRC}}^{\dagger}\tilde{\bm{w}}\right\Vert ^{2}|\bm{H}\})P(\bm{x}|\mathcal{H}_{i})=\\
\tilde{\bm{z}}_{{\scriptscriptstyle MRC}}^{\dagger}\tilde{\bm{H}}\bm{C}(\bm{x}|\mathcal{H}_{i})\tilde{\bm{H}}^{\dagger}\tilde{\bm{z}}_{{\scriptscriptstyle MRC}}+\frac{\sigma_{w}^{2}}{2}\left\Vert \bm{z}_{{\scriptscriptstyle MRC}}\right\Vert ^{2},
\end{gather}
}where $\bm{C}(\bm{x}|\mathcal{H}_{i})\triangleq\mathbb{E}\{(\bm{x}-\mathbb{E}\{\bm{x}|\mathcal{H}_{i}\})(\bm{x}-\mathbb{E}\{\bm{x}|\mathcal{H}_{i}\})^{T}|\mathcal{H}_{i}\}$.
Therefore, in view of these results and exploiting Eq. (\ref{eq:GMM_MRC}),
we obtain the following low-$\mathrm{SNR}$ approximation for $P_{F_{0}}(\bm{H})$:

\begin{equation}
P_{F_{0}}(\bm{H})\approx\mathcal{Q}\left(\frac{\gamma-\Re(\bm{z}_{{\scriptscriptstyle MRC}}^{\dagger}\bm{H}\mathbb{E}\{\bm{x}|\mathcal{H}_{0}\})}{\sqrt{\tilde{\bm{z}}_{{\scriptscriptstyle MRC}}^{\dagger}\tilde{\bm{H}}\bm{C}(\bm{x}|\mathcal{H}_{0})\tilde{\bm{H}}^{\dagger}\tilde{\bm{z}}_{{\scriptscriptstyle MRC}}+\frac{\sigma_{w}^{2}}{2}\left\Vert \bm{z}_{{\scriptscriptstyle MRC}}\right\Vert ^{2}}}\right).\label{eq:IC_PF0_lowSNR}
\end{equation}
Under the simplifying assumptions $\mathbb{E}\{\bm{x}|\mathcal{H}_{0}\}=(2P_{F}-1)\bm{1}_{K}$
and $\bm{C}(\bm{x}|\mathcal{H}_{0})=\left[1-(2P_{F}-1)^{2}\right]\bm{I}_{K}$
(i.e. the threshold of each sensor is set to assure the same $P_{F}$
and also the decisions are uncorrelated, given $\mathcal{H}_{0}$),
Eq. (\ref{eq:IC_PF0_lowSNR}) simplifies to:
\begin{equation}
P_{F_{0}}(\bm{H})\approx\mathcal{Q}\left(\frac{\gamma-\delta\cdot\left\Vert \bm{z}_{{\scriptscriptstyle MRC}}\right\Vert ^{2}}{\sqrt{(1-\delta^{2})\cdot\tilde{\bm{z}}_{{\scriptscriptstyle MRC}}^{\dagger}\tilde{\bm{H}}\tilde{\bm{H}}^{\dagger}\tilde{\bm{z}}_{{\scriptscriptstyle MRC}}+\frac{\sigma_{w}^{2}}{2}\left\Vert \bm{z}_{{\scriptscriptstyle MRC}}\right\Vert ^{2}}}\right),\label{eq:ICPF0_lowSNR_uncorr}
\end{equation}
where $\delta\triangleq(2P_{F}-1)$. Eq. (\ref{eq:ICPF0_lowSNR_uncorr})
still contains a problematic dependence w.r.t. the entire channel
matrix $\bm{H}$; therefore we consider a large-system ($K\rightarrow+\infty)$
regime, where $\tilde{\bm{H}}\tilde{\bm{H}}^{\dagger}\approx2K\bm{I}_{N}$
holds, thus leading to
\begin{equation}
P_{F_{0}}(\bm{H})\approx\mathcal{Q}\left(\frac{\gamma-\delta\left\Vert \bm{z}_{{\scriptscriptstyle MRC}}\right\Vert ^{2}}{\sqrt{\frac{1}{2}\left((1-\delta^{2})\cdot K+\sigma_{w}^{2}\right)}\left\Vert \bm{z}_{{\scriptscriptstyle MRC}}\right\Vert }\right),
\end{equation}
which can be easily inverted to provide Eq. (\ref{eq:low-SNR large-system threshold}).

\section{Proof of Proposition \ref{prop:MGF_mrc}\label{sec:Appendix_MGF-MRC}}

We derive here the closed form of $\Phi_{-\Lambda_{MRC}}(s|\bm{x})$
in Eq. (\ref{eq:mgf_lambda_mrc}). Similarly to \cite{Zhu2002}, where
the symbol-error probability in a fading environment with antenna
diversity was obtained, we express $-\Lambda_{{\scriptscriptstyle MRC}}|\bm{x}$
as follows
\begin{equation}
-\Lambda_{{\scriptscriptstyle MRC}}|\bm{x}=\sum_{n=1}^{N}(\bm{v}_{n}|\bm{x})^{\dagger}\cdot\bm{F}\cdot(\bm{v}_{n}|\bm{x}),\label{eq:lambda_mrc_quadform}
\end{equation}
where the Gaussian vectors $(\bm{v}_{n}|\bm{x})$, $n\in\{1,\ldots,N\}$,
and the deterministic matrix $\bm{F}$ have the explicit expressions:
\begin{align}
\bm{v}_{n}|\bm{x} & \triangleq\left[\begin{array}{cc}
y_{n}|\bm{x} & \bm{h}_{r,n}\bm{1}_{K}\end{array}\right]^{t}\qquad\bm{F}\triangleq\left[\begin{array}{cc}
0 & -\frac{1}{2}\\
-\frac{1}{2} & 0
\end{array}\right]\label{eq:vn and F}
\end{align}
with $\bm{h}_{r,n}$ denoting the $n$th row of $\bm{H}$. Note that
Eq. (\ref{eq:lambda_mrc_quadform}) is a sum of Hermitian quadratic
forms of circularly complex Gaussian vectors $\bm{v}_{n}|\bm{x}$.
Since $\bm{v}_{n}|\bm{x}$ are i.i.d. vectors, the Laplace transform
of Eq. (\ref{eq:lambda_mrc_quadform}) has the following closed form
\cite{Schwarz1966}:
\begin{equation}
\Phi_{-\Lambda_{MRC}}(s|\bm{x})=\left[\frac{1}{\det(\bm{I}_{2}+s\bm{L}(\bm{x}))}\right]^{N}
\end{equation}
where $\bm{L}(\bm{x})\triangleq(\bm{R}|\bm{x})\cdot\bm{F}$ and $\bm{R}|\bm{x}\triangleq\mathbb{E}\left\{ (\bm{v}_{n}|\bm{x})(\bm{v}_{n}|\bm{x})^{\dagger}\right\} $,
i.e. the covariance matrix of $\bm{v}_{n}|\bm{x}$, since $\mathbb{E}\left\{ (\bm{v}_{n}|\bm{x})\right\} =\bm{0}_{2}$.
The explicit expression of $\bm{R}|\bm{x}$ is: 
\begin{eqnarray}
\bm{R}|\bm{x} & = & \mathbb{E}\left\{ (\bm{v}_{n}|\bm{x})(\bm{v}_{n}|\bm{x})^{\dagger}\right\} \\
 & = & \left[\begin{array}{cc}
K+\sigma_{w}^{2} & 2\ell(\bm{x})-K\\
2\ell(\bm{x})-K & K
\end{array}\right].\label{eq:R|x}
\end{eqnarray}

Denoting $\lambda_{i}(\bm{x})$, $i\in\{1,2\}$, the two eigenvalues
of $\bm{L}(\bm{x})$ we have that
\begin{align}
\Phi_{-\Lambda_{MRC}}(s|\bm{x}) & =\left[\frac{1}{(1+s\lambda_{1}(\bm{x}))(1+s\lambda_{2}(\bm{x}))}\right]^{N}.\label{eq:lt_mrc}
\end{align}
Evaluation of $\lambda_{i}(\bm{x})$, through $\det(s\bm{I}_{2}-\bm{L}(\bm{x}))=0$,
gives: 
\begin{equation}
\lambda_{i}(\bm{x})=\frac{1}{2}\cdot\left(K-2\ell(\bm{x})\pm K\sqrt{1+\frac{1}{\mathrm{SNR}}}\right),\label{eq:eig_mrc}
\end{equation}
where we have exploited that $\mathrm{SNR}=\frac{K}{\sigma_{w}^{2}}$.
Direct substitution of explicit expression of $\lambda_{i}(\bm{x})$
in Eq. (\ref{eq:lt_mrc}) provides the result.

\section{Proof of Proposition \ref{prop:largesystemMGFs}\label{sec:Appendix_LargesystemMGF}}

In this Appendix we prove the large system conditional MGFs given
by Eqs. (\ref{eq:MGFH1_kinf_IPC}) and (\ref{eq:MGFH0_kinf_IPC})
and by Eqs. (\ref{eq:MGFH1_kinf_TPC}) and (\ref{eq:MGFH0_kinf_TPC})
in the IPC and TPC scenarios, respectively. We show the proof for
the IPC case; differences with the TPC scenario will be underlined
throughout the demonstration. We start by giving the definitions
\begin{align}
\bm{p}_{n}\triangleq\frac{1}{\sqrt{K}}\left[\begin{array}{cc}
y_{n} & \bm{h}_{r,n}\bm{1}_{K}\end{array}\right]^{t},\qquad & \bm{p}\triangleq\left[\begin{array}{ccc}
\bm{p}_{1}^{t} & \cdots & \bm{p}_{N}^{t}\end{array}\right]^{t},
\end{align}
with $\bm{h}_{r,n}$ still denoting the $n$th row of $\bm{H}$. It
can be noticed that $\bm{p}_{n}|\bm{x}=\frac{1}{\sqrt{K}}\bm{v}_{n}|\bm{x}$,
where $\bm{v}_{n}|\bm{x}$ has been defined in Eq. (\ref{eq:vn and F}).
Based on this observation, we can conclude that ($i$) $\bm{p}_{n}|\bm{x}$,
$n\in\{1,\ldots,N\}$, are i.i.d. circularly complex Gaussian vectors
($ii$) $\mathbb{E}\{\bm{p}_{n}|\bm{x}\}=\frac{1}{\sqrt{K}}\mathbb{E}\{\bm{v}_{n}|\bm{x}\}=\bm{0}_{2}$
and $\breve{\bm{R}}|\bm{x}\triangleq\mathbb{E}\{\bm{p}_{n}\,\bm{p}_{n}^{\dagger}|\bm{x}\}=\frac{1}{K}\bm{R}|\bm{x}$
(cf. Eq. (\ref{eq:R|x})) and finally ($iii$) $\bm{p}|\bm{x}$ is
a circularly complex Gaussian vector, whose \emph{characteristic function
}(CF), denoted $\Omega(\cdot)$, can be expressed as a function of
the dual vectors $\bar{\bm{s}}_{n}\triangleq\left[\begin{array}{cc}
s_{1,n} & s_{2,n}\end{array}\right]^{t}\in\mathbb{C}^{2}$, $n\in\{1,\ldots N\}$, as follows \cite{Picinbono1996}:
\begin{equation}
\Omega_{\bm{p}}(\bar{\bm{s}}|\bm{x})=\exp\left[-\frac{1}{4}\sum_{n=1}^{N}\bar{\bm{s}}_{n}^{\dagger}\cdot\left(\breve{\bm{R}}|\bm{x}\right)\cdot\bar{\bm{s}}_{n}\right].\label{eq:CF|x_compact}
\end{equation}
By exploiting the structure of $\breve{\bm{R}}|\bm{x}$, we can expand
$\Omega_{\bm{p}}(\bar{\bm{s}}|\bm{x})$ as follows
\begin{gather}
\Omega_{\bm{p}}(\bar{\bm{s}}|\bm{x})=\exp\left[-\frac{1}{K}\sum_{n=1}^{N}\Re\{s_{1,n}^{*}s_{2,n}\}\right]^{\ell(\bm{x})}\times\label{eq:CF|x}\\
\exp\left[-\frac{1}{4}\sum_{n=1}^{N}\left(\left\Vert s_{1,n}\right\Vert ^{2}\cdot(1+\frac{\sigma_{w}^{2}}{K})+\left\Vert s_{2,n}\right\Vert ^{2}-2\Re\{s_{1,n}^{*}s_{2,n}\}\right)\right].\nonumber 
\end{gather}
W.l.o.g. we focus hereinafter on $\Omega_{\bm{p}}(\bar{\bm{s}}|\mathcal{H}_{1})=\sum_{\ell(\bm{x})=0}^{K}\Omega_{\bm{p}}(\bar{\bm{s}}|\ell(\bm{x}))P(\ell(\bm{x})|\mathcal{H}_{1})$
(since identical considerations apply to $\Omega_{\bm{p}}(\bar{\bm{s}}|\mathcal{H}_{0})$)
and we recall that for the conditional i.i.d. sensor decisions $\ell(\bm{x})|\mathcal{H}_{1}\sim\mathcal{B}(K,P_{D})$.
Exploiting this assumption and Eq. (\ref{eq:CF|x}), we get the explicit
expression (we drop the dependence of $\ell$ w.r.t. $\bm{x}$):
\begin{gather}
\Omega_{\bm{p}}(\bar{\bm{s}}|\mathcal{H}_{1})=\sum_{\ell=0}^{K}\left(\begin{array}{c}
K\\
\ell
\end{array}\right)P_{D}^{\ell}(1-P_{D})^{K-\ell}\times\nonumber \\
\exp\left[-\frac{1}{K}\sum_{n=1}^{N}\Re\{s_{1,n}^{*}s_{2,n}\}\right]^{\ell}\times\nonumber \\
\exp\left[-\frac{1}{4}\sum_{n=1}^{N}\left(\left\Vert s_{1,n}\right\Vert ^{2}\cdot(1+\frac{\sigma_{w}^{2}}{K})+\left\Vert s_{2,n}\right\Vert ^{2}-2\Re\{s_{1,n}^{*}s_{2,n}\}\right)\right]\\
=\left((1-P_{D})+P_{D}\exp\left[-\frac{1}{K}\sum_{n=1}^{N}\Re\{s_{1,n}^{*}s_{2,n}\}\right]\right)^{K}\times\nonumber \\
\exp\left[-\frac{1}{4}\sum_{n=1}^{N}\left(\left\Vert s_{1,n}\right\Vert ^{2}\cdot(1+\frac{\sigma_{w}^{2}}{K})+\left\Vert s_{2,n}\right\Vert ^{2}-2\Re\{s_{1,n}^{*}s_{2,n}\}\right)\right]\label{eq:expanded CF|H1}
\end{gather}
Also, using table of limits, eventually we have that:
\begin{gather}
\bar{\Omega}_{\bm{p}}(\bar{\bm{s}}|\mathcal{H}_{1})\triangleq\lim_{K\rightarrow+\infty}\Omega_{\bm{p}}(\bar{\bm{s}}|\mathcal{H}_{1})\\
=\exp\left(-P_{D}\sum_{n=1}^{N}\Re\{s_{1,n}^{*}s_{2,n}\}\right)\times\nonumber \\
\exp\left[-\frac{1}{4}\sum_{n=1}^{N}\left(\left\Vert \bar{\bm{s}}_{n}\right\Vert ^{2}-2\Re\{s_{1,n}^{*}s_{2,n}\}\right)\right]\\
=\exp\left[-\frac{1}{4}\sum_{n=1}^{N}\left(\left\Vert \bar{\bm{s}}_{n}\right\Vert ^{2}+2\cdot(2P_{D}-1)\cdot\Re\{s_{1,n}^{*}s_{2,n}\}\right)\right].\label{eq:ccf to be casted}
\end{gather}
It is worth noticing that $\bar{\Omega}_{\bm{p}}(\bar{\bm{s}}|\mathcal{H}_{1})$
in TPC scenario is obtained by setting $\frac{1}{\mathrm{SNR}}=\frac{\sigma_{w}^{2}}{K}$
in Eq. (\ref{eq:expanded CF|H1}) and evaluating $\lim_{K\rightarrow+\infty}\Omega_{\bm{p}}(\bar{\bm{s}}|\mathcal{H}_{1})$
analogously. The expression in Eq. (\ref{eq:ccf to be casted}) can
be recast as:
\begin{eqnarray}
\bar{\Omega}_{\bm{p}}(\bar{\bm{s}}|\mathcal{H}_{1}) & = & \exp\left[-\sum_{n=1}^{N}\frac{1}{4}\bar{\bm{s}}_{n}^{\dagger}\cdot\bar{\bm{R}}\cdot\bar{\bm{s}}_{n}\right];\\
\bar{\bm{R}} & \triangleq & \left[\begin{array}{cc}
1 & 2P_{D}-1\\
2P_{D}-1 & 1
\end{array}\right].
\end{eqnarray}
Such a result, when compared with Eq. (\ref{eq:CF|x_compact}) and
with the use of \emph{Levi's continuity theorem }\cite{Karr1993},
states that when $K\rightarrow+\infty$, $\bm{p}_{n}|\mathcal{H}_{1}$,
$n\in\{1,\ldots,N\}$, are i.i.d and $\bm{p}_{n}|\mathcal{H}_{1}\overset{d}{\rightarrow}\mathcal{N}_{\mathbb{C}}(\bm{0}_{2},\bar{\bm{R}})$.
Analogously, in the TPC scenario a similar result holds when $\bar{\bm{R}}$
is appropriately replaced. Finally, this information is readily exploited
by considering that $\tilde{\Lambda}\triangleq\frac{1}{K}\Lambda_{{\scriptscriptstyle MRC}}$
equals to
\begin{equation}
-\tilde{\Lambda}=\sum_{n=1}^{N}\bm{p}_{n}^{\dagger}\cdot\bm{F}\cdot\bm{p}_{n},
\end{equation}
where $\bm{F}$ has the same definition as in Eq. (\ref{eq:vn and F}).
Therefore, as $K\rightarrow+\infty$, $-\tilde{\Lambda}|\mathcal{H}_{1}$
is a sum of Hermitian quadratic forms of i.i.d. circularly complex
Gaussian vectors whose MGF is easily derived using similar arguments
as in Appendix \ref{sec:Appendix_MGF-MRC}, thus providing Eqs. (\ref{eq:MGFH1_kinf_IPC})
and (\ref{eq:MGFH0_kinf_IPC}) (and analogously Eqs. (\ref{eq:MGFH1_kinf_TPC})
and (\ref{eq:MGFH0_kinf_TPC})).

\section{Proof of Proposition \ref{prop:MGF_AUC}\label{sec:Appendix_MGF_AUC}}

The first step in proving Eq. (\ref{eq:AUC_Laplace Domain}) is showing
that, after some manipulations, the $\mathrm{AUC}$ defined in Eq.
(\ref{eq:AUC_main_formula}) for a generic statistic $\Lambda$ can
be expressed in the alternative form:
\begin{equation}
\mathrm{AUC}=\intop_{-\infty}^{+\infty}P_{D_{0}}(\gamma)p_{-\Lambda}(-\gamma|\mathcal{H}_{0})d\gamma\label{eq:AUC_manipulated_form}
\end{equation}
 where $p_{-\Lambda}(\lambda|\mathcal{H}_{i})$, $\mathcal{H}_{i}\in\mathcal{H}$,
denotes the conditional pdf of $-\Lambda$. Furthermore, it can be
shown that the Laplace transforms of $P_{D_{0}}(\gamma)$ and $p_{-\Lambda}(-\gamma|\mathcal{H}_{0})$
are given by $\frac{\Phi_{-\Lambda}(-s|\mathcal{H}_{1})}{-s}$ and
$\Phi_{-\Lambda}(-s|\mathcal{H}_{0})$, respectively. Also, let us
recall the relationship between a generic function $g(\lambda)$ and
its two-sided Laplace transform $\Phi_{g}(s)\triangleq\int_{-\infty}^{+\infty}g(\lambda)\exp\left(-\lambda s\right)ds$
\begin{eqnarray}
\lim_{s\rightarrow0}\Phi_{g}(s) & = & \int_{-\infty}^{+\infty}g(\lambda)d\lambda,\label{eq:Laplace integration-1}
\end{eqnarray}
and the property relating the Laplace transform of the product of
two generic functions $a(\lambda)$ and $b(\lambda)$ 
\begin{equation}
\Phi_{a\cdot b}(s)=\frac{1}{2\pi j}\intop_{\alpha-j\infty}^{\alpha+j\infty}\Phi_{a}(p)\Phi_{b}(s-p)dp.\label{eq:Laplace convolution-1}
\end{equation}
where $\alpha$ is a constant that ensures that the integration is
performed in the RC of $\Phi_{a}(p)$. Combining Eqs. (\ref{eq:Laplace integration-1})
and (\ref{eq:Laplace convolution-1}) we get 
\begin{equation}
\frac{1}{2\pi j}\intop_{\alpha-j\infty}^{\alpha+j\infty}\Phi_{a}(p)\Phi_{b}(-p)dp=\int_{-\infty}^{+\infty}a(\lambda)b(\lambda)d\lambda.\label{eq:Innerproduct_Laplacedomain-1}
\end{equation}
The obtained expression is now used to evaluate Eq. (\ref{eq:AUC_manipulated_form})
in the Laplace domain. In fact, exploiting the explicit expressions
of the Laplace transforms of $P_{D_{0}}(\gamma)$ and $p_{-\Lambda}(-\gamma|\mathcal{H}_{0})$
in Eq. (\ref{eq:Innerproduct_Laplacedomain-1}) we obtain
\begin{equation}
\mathrm{AUC}=\frac{1}{2\pi j}\intop_{\alpha-j\infty}^{\alpha+j\infty}\frac{\Phi_{-\Lambda}(-p|\mathcal{H}_{1})}{-p}\Phi_{-\Lambda}(p|\mathcal{H}_{0})dp,\label{eq:AUC_MGF_appendix_inverted_sign}
\end{equation}
where we can choose $\alpha=-c_{1}$, where $c_{1}$ has the same
meaning as in Eq. (\ref{eq:laplace_integral}), i.e. belongs to the
positive restriction of the RC of $\Phi_{-\Lambda}(s|\mathcal{H}_{1})$.
Finally, the substitution $s^{*}=-p$ in Eq. (\ref{eq:AUC_MGF_appendix_inverted_sign})
gives the result in Eq. (\ref{eq:AUC_Laplace Domain}).

\bibliographystyle{IEEEtran}
\bibliography{IEEEabrv,sensor_networks}

\end{document}